\documentclass[11pt]{article}
\usepackage{amsfonts}
\usepackage{amsbsy} 
\usepackage{epsfig}
\usepackage{latexsym}
\usepackage{graphicx}
\usepackage{amsmath, mathrsfs, xcolor}
\usepackage[labelformat=simple]{subcaption}

\usepackage{breqn}
\usepackage[style=numeric-comp, backref=true, sorting=none, url=false, eprint=false, maxbibnames=9]{biblatex}
\usepackage{hyperref}
\hypersetup{hidelinks}
\usepackage{setspace, soul}
\usepackage[right]{lineno}
\usepackage[margin=0.8in]{geometry}
\allowdisplaybreaks

\usepackage{listings} 
\makeatletter
\@addtoreset{equation}{section}
\makeatother

\begin{document}
\hfuzz=10pt
\title{{\Large \bf{Spacetime Structure of a Regular Accelerating Black Hole Pair in General Relativity}}}
\author{M M Akbar\footnote{akbar@utdallas.edu} \footnote{Author to whom any correspondence should be addressed.} $^a$
\ \,\,\&
C P Brewer\footnote{charlie.brewer@utdallas.edu} $^a$
\ \,\,\&
S M Modumudi\footnote{saimadhav.modumudi@utdallas.edu} $^b$
\\
\\
$^a$ Department of Mathematical Sciences,
\\ University of Texas at Dallas, 
\\ 800 W Campbell Rd,
\\ Richardson, Texas, USA.
\\
\\
$^b$ Department of Physics,
\\ University of Texas at Dallas, 
\\ 800 W Campbell Rd,
\\ Richardson, Texas, USA.}
\date{\today}
\maketitle
\begin{abstract}
We revisit the one-parameter generalization of the C-metric derived by Ernst, which solves the vacuum Einstein equations. Resolving conflicting claims in the literature, we determine the correct value of the parameter that ensures the regularity of the metric on the axis. This ``regularized C-metric" describes a pair of accelerating black holes without the line source present in the original C-metric. Additionally, this generalization changes the Petrov type from D to I.  We use the Gauss-Bonnet theorem to analyze the nodal singularities, the line source, and their relation to the horizon topology. Both the black hole and acceleration horizons are found to be embeddable in $\mathrm{E}^3$. We examine various geometric and asymptotic properties in detail using several coordinate systems and construct the corresponding 2D and 3D conformal diagrams. This process is more involved than for the original C-metric due to the presence of the exponential factors. These exponential factors also introduce curvature singularities at infinity, which obstructs asymptotic flatness. Contrary to Bonnor's expectation, we demonstrate why Bondi’s algorithm for obtaining the standard Bondi form fails for the C-metric, despite its asymptotic flatness. We also show that Ernst's solution-generating prescription in boost-rotation symmetric coordinates is a symmetry of the wave equation.
\end{abstract}


\section{Introduction}
The C-metric is a vacuum solution of Einstein's equations, first discovered by Levi-Civita in 1918 and independently a year later by Weyl in its static form \cite{LC, Weyl_C-metric}. It has been rediscovered multiple times since and was given this curious name by Ehlers and Kundt \cite{Newman_Tamburino, Robinson_Trautman, ehlers_kundt_review}. Robinson and Trautman were the first to identify the $r^{-1}$ fall-off of its Riemann tensor, which is characteristic of radiation, and interpreted the C-metric as simultaneously static and radiative \cite{Robinson_Trautman}. It took more than half a century to realize that the maximal extension of this spacetime describes two causally separated black holes accelerating away from each other and that the conical singularities in the metric represent a strut or a string causing the acceleration \cite{Kinnersley_Walker, Bonnor}. This picture remains valid for the charged and other generalizations of the C-metric. Following this realization, C-metrics became the subject of intense interest, and their geometric and asymptotic structures were extensively studied (see, for example, \cite{griffiths_krtous_podolsky, griffiths_podolsky_2009, DiasLemos} and references therein). In the 1990s, interest in C-metrics acquired a new dimension in connection with their possible role in pair creation of black holes in quantum gravity (see, for example, \cite{Ross-2022}, and references therein). More recently, they have inspired a new line of research into the thermodynamics of accelerating black holes with conical singularities in which they partake in the first law \cite{AppelsGregory_16, AppelsGregory_17}. There has been persistent interest in C-metrics as exact axially symmetric local black hole models. In particular, geodesics, lensing effects, shadows, (anti)-photon surfaces have been analyzed in considerable detail \cite{BicakRees, Pravda_2001, Bini_2005, Bini:2007, Alawadi_2020, Batic_2021, Grenzebach_2015, Batic_2021, Grenzebach_2015, GibbonsWarnick}.

How to deal with the conical singularities of the C-metric has long been a central issue. They are interpreted as cosmic strings or Weyl struts (see Section \ref{con-sing-sec}). Recent work shows that they can be seen as null dust associated with the momentum flux of gravitational radiation \cite{Kofron:2020}. To remove the conical or ``nodal" singularities of the charged C-metric, Ernst ``immersed" the metric in an external electric field: he applied a Harrison-type transformation and obtained a one-parameter generalization, the famous ``Ernst solution" \cite{Ernst:1976}. For a certain value of the parameter, the resulting Einstein-Maxwell solution is free from conical singularities. Ernst also immersed the vacuum C-metric in an external gravitational field by inventing what is now known as his eponymous solution-generating technique \cite{Ernst1978, Ernst1979}. He expected that this one-parameter vacuum generalization would be free from conical singularities for a specific parameter value and conjectured that this ``generalized C-metric would serve as a better model for an accelerating black hole than the C-metric itself, since a physical origin of the acceleration will have been provided." Using a linear approximation, Hogan argued that the entire axis could not be regularized for this metric, except at a single arbitrary value of the radial coordinate \cite{Hogan_1979}. Dray and Walker subsequently showed, without relying on approximations, that the metric can indeed be regularized; their parameter value, however, differed from the one conjectured by Ernst \cite{TDray_MWalker}. Later, Bonnor, seemingly unaware of Dray and Walker’s work, derived this generalized C-metric in Weyl coordinates, which revealed the Newtonian rod sources of the metric \cite{Bonnor88}. He obtained yet another value of the parameter that regularizes the conical singularities. We will revisit these works in the course of this paper. Another interpretation of the generalized vacuum C-metric comes from the work of Bi\v{c}\'ak, Hoenselaers, and Schmidt, who obtained it as a limiting case of the Curzon-Chazy metrics, where one particle in each pair is moved to infinity while simultaneously increasing their mass \cite{BHS-1, BHS-2}.

Apart from the works mentioned above, the regularized vacuum C-metric has not been studied in detail as a solution. Instead, more attention has been accorded to non-vacuum generalizations of the C-metric. In this paper, we will address this gap. We will first work out the correct parameter values for regularization and obtain the metric in different coordinate systems. This will,  a fortiori, lead us to revisit the original C-metric in various coordinate systems, make some of its properties more salient, and make some novel observations. 

Ernst's generalization of the C-metric can be seen as a special case of a wider class of spacetimes obtained recently via the inverse scattering method in Weyl coordinates  \cite{Astorino21, BHS-2}. While Ernst's prescription can be implemented for a wider class of coordinates than just Weyl coordinates, it is not immediately clear how to do the same for this larger class of solutions. We hope that this work will serve as a precursor to such future investigations.

The rest of the paper is organized as follows: in Section \ref{c-intro}, we review the C-metric, its conical singularities, and their interpretation, and connect the discussion to the Gauss-Bonnet theorem. In Section \ref{ernst-c-metric}, we derive the generalized C-metric in Hong-Teo coordinates and discuss its regularization and geometric properties. In Section \ref{linesource}, we analyze the effect of Ernst's prescription on the line sources in the sense of Israel using Weyl coordinates. We also examine the boost-rotation symmetric form of the metric, reconfirming that the interpretation as a pair of accelerating black holes remains valid. Finally, in Section \ref{asym_str}, we show that the spacetime is not asymptotically flat at null infinity, and obtain 2D and 3D conformal diagrams.

\section{The C-metric}\label{c-intro}
Following the work of Kinnersley and Walker
\cite{Kinnersley_Walker}, the most widely used form of the C-metric is most likely the following:
\begin{equation}
\label{kinnersly-walker}
ds^2  = \frac{1}{a^2 (\tilde{x}+\tilde{y})^2}\left[-\tilde{A} d\tilde{t}^2 + \frac{1}{\tilde{A}}d\tilde{y}^2 + \frac{1}{\tilde{B}}d\tilde{x}^2  + \tilde{B} d\tilde{\phi}^2\right]~,
\end{equation}
where
\begin{equation}
\tilde{A} = -1 + \tilde{y}^2 - 2Ma\tilde{y}^3~, \qquad  \tilde{B} = 1 - \tilde{x}^2 - 2Ma\tilde{x}^3~, \qquad \text{with} \qquad  -\infty<\tilde{x},\tilde{y}<\infty~.
\end{equation}
$\tilde{A}$ and $\tilde{B}$ will have three real roots each if $27M^2a^2<1$, which will be expressions in terms of $M$ and $a$. As shown by Hong and Teo in 2003, with a redefinition of coordinates and parameters ($M$ and $a$) the metric can be written as \cite{Hong_Teo} (see Appendix \ref{Dray-Walker-connection})
\begin{equation}\label{C-metric-HongTeo}
ds^2  = \frac{1}{\alpha^2(x+y)^2}\left[-Ad\tau^2 + \frac{1}{A}dy^2 + \frac{1}{B}dx^2  + B d\varphi^2\right]~,
\end{equation}
where
\begin{equation}\label{C-metric-HongTeo-func}
A = -(1-y^2)(1-2\alpha my)~, \qquad B = (1-x^2)(1 + 2\alpha mx)~.
\end{equation}
This makes the roots appear at $\pm 1,$ and $\pm 1/2\alpha m$.  The condition $0<2\alpha m<1$ preserves the order of roots. The metric (\ref{C-metric-HongTeo}) is static for $A\cdot B>0$ and is of Petrov type D (or $\{2,2\}$) with the only nonvanishing Weyl scalar $\Psi_2 \equiv -m\alpha^3 (x+y)^3$. The coordinates $x,y$ must be restricted such that $x+y$ is strictly positive (or negative) for the metric to represent a physical spacetime since $x+y=0$ is conformal infinity. With the choice $A, B>0$, the coordinates are restricted to $-1 \leq x \leq 1$ and $-x < y < \infty$ (shaded region in Figure \ref{CoordPlot}). This part of the spacetime is generally referred to as the ``physical" spacetime, which is what will concern us in this paper\footnote{The remaining coordinate ranges correspond to distinct spacetimes, see \cite{pravda-pravdova, Sladek-Finley}.}. For generality, one takes $\varphi\in[0,2\pi C)$, $C>0$. The static part of the spacetime is bounded by two Killing horizons, that is, the black hole horizon at $y=1/2\alpha m$ and the acceleration horizon at $y=1$. Kinnersley and Walker also wrote the metric using retarded null coordinates and a radial coordinate
\begin{equation}
\label{c-metric-null}
    \alpha u = \tau + \int A^{-1} dy, \hspace{2em} \alpha r = (x+y)^{-1}
\end{equation}
which puts the metric in the form
\begin{equation}
\label{null-coordinates}
    ds^2 = H du^2 + 2du dr + 2\alpha r^2 du dx  -r^2 (B^{-1}dx^2 + Bd\phi^2)
\end{equation}
where 
\begin{equation}
    H=-\alpha^2 r^2  \left(x-\frac{1}{\alpha r}\right){B}.
\end{equation}
This allows the spacetime to be extended beyond either horizon like the Kruskal extension of the Schwarzschild metric. Setting $x=\cos\theta$, one can cast the metric into a more familiar form:
\begin{equation}
    ds^2 = H du^2 + 2du dr - 2\alpha r^2 \sin\theta du d\theta - r^2(B^{-1} \sin^2\theta d\theta^2 + Bd\phi^2). 
\end{equation}

\subsubsection*{Spherical{ -like} Coordinates}
Since $-1 \leq x \leq 1$, one can write  (\ref{C-metric-HongTeo})- (\ref{C-metric-HongTeo-func}) in spherical-like coordinates \cite{ Griffiths-Podolsky:2005, griffiths_krtous_podolsky, Griffiths-Podolsky:2005, griffiths_podolsky_2009}\footnote{We use $x = -\cos\theta$ instead of $x = \cos\theta$ since then it is easier to keep the signs consistent.}$^,$\footnote{This (spherical) $r$ is different than  $r$ used in equation \eqref{c-metric-null}}
\begin{equation}\label{HT-sph-trans}
x = -\cos\theta~, \quad y = \frac{1}{\alpha r}~, \quad \tau = \alpha t~,
\end{equation}
making the metric
\begin{equation}\label{c-metric-metric}
ds^2 = \frac{1}{(1-\alpha r \cos\theta)^2}\left[ -Qdt^2 + \frac{1}{Q}dr^2 + \frac{r^2}{P}d\theta^2 + P r^2 \sin^2\theta d\varphi^2\right]~,
\end{equation}
where
\begin{equation}\label{c-metric-func}
Q = \left( 1- \frac{2m}{r} \right)\left(1-\alpha^2 r^2\right)~, \qquad P = 1 - 2\alpha m \cos\theta~, \qquad \text{and} \qquad  0<r<\frac{1}{\alpha\cos\theta}~.
\end{equation}
With this form, one recovers the Schwarzschild metric and the Rindler spacetime for $\alpha = 0$ and $m = 0$, respectively. Thus, $m$ and $\alpha$ are referred to as the mass and acceleration parameters of the metric. The black hole and the acceleration horizons are located at $r=2m$ and $r=1/\alpha$, respectively. Thus, in the static region, $2m < r< 1/\alpha$,   the metric \eqref{c-metric-metric} -- \eqref{c-metric-func} is a local axisymmetric black hole spacetime in the sense of Geroch, Hartle, and Chandrasekhar \cite{GerochHartle, chandrasekharbook}. Taking the limit $m=0$, and by using appropriate coordinate transformations, one can interpret that the black hole accelerates with respect to a distant inertial observer following Newton's laws \cite{BicakRees}. Note that the radial coordinate does not cover the $y\in (-1, 0]$ part of the ``physical" spacetime, for which one needs to revert to the $x,y$ coordinates.

\subsubsection*{Curvature Singularities}
For a four-dimensional spacetime, there are fourteen independent algebraic invariants of the Riemann curvature tensor \cite{ZakharyMcIntosh}. For a vacuum spacetime this number reduces to four, and in the case of C-metric, only two of them are non-zero. For \eqref{C-metric-HongTeo} - \eqref{C-metric-HongTeo-func}, they are given by\footnote{The more familiar curvature scalars $I$ and $J$ are given by $I = \frac{1}{2} w_1$, and $J = \frac{1}{6} w_2$. Note that the Weyl scalar $\Psi_2$ is given by $\Psi_2 = -\sqrt{\frac{w_1}{6}}$.}
\begin{equation} \label{eq:curvaturescalars}
    w_1 = 6 m^2 \alpha^6 (x+y)^6 \qquad \text{and} \qquad w_2 = 6 m^3 \alpha^9 (x+y)^9~,
\end{equation}
which in spherical-like coordinates become \eqref{c-metric-metric} - \eqref{c-metric-func}
\begin{equation}
    w_1  = \frac{6 \left(1 - \alpha  r \cos\theta\right)^{6} m^{2}}{r^{6}} \qquad \text{and} \qquad w_2 = \frac{6 \left(1 - \alpha  r \cos\theta \right)^{9} m^{3}}{r^{9}}~.
\end{equation}
This clearly shows that the only singularities are at $r=0$. However, the C-metric has conical singularities that shall be addressed in Section \ref{con-sing-sec}.

\subsubsection*{Weyl Coordinates}
As with any static axisymmetric vacuum metric, the static part of the C-metric $(2m<r<1/\alpha)$, which we will call region II in subsequent discussions, can be written as a Weyl solution
\begin{equation}\label{weyl-metric}
ds^2 = -e^{2u}dt^2 + e^{-2u+2\nu}(d\rho^2 + dz^2) + \rho^2 e^{-2u} d\varphi^2~,
\end{equation}
where the functions $u(\rho,z),~\nu(\rho,z)$ satisfy the differential equations (see, for example, \cite{stephani_exact_sol})
\begin{equation}
u_{,\rho\rho} + \frac{1}{\rho}u_{,\rho} + u_{,zz} = 0~, \qquad  \nu_{,\rho} = \rho\left( u_{,\rho}^2 - u_{,z}^2 \right) \qquad \text{and} \qquad \nu_{,z} = 2 u_{,\rho} u_{,z}~.
\end{equation}
For the C-metric, these functions are given by (see, for example, \cite{griffiths_podolsky_2009})
\begin{align}
\label{weyl-u} e^{2u} &= \alpha \frac{R_1+R_2-2m}{R_1+R_2+2m} \left( R_3 + z + \frac{1}{2\alpha} \right)~,\\
e^{-2u+2\nu} &= \frac{\left[ (1-2\alpha m)R_1 + (1+2\alpha m)R_2 + 4\alpha m R_3 \right]^2}{8\alpha \left(1-4\alpha^2 m^2\right)^2 R_1 R_2 R_3}~,
\end{align}
with
\begin{equation}
    R_i \equiv \sqrt{\rho^2 + (z-z_i)^2}~, \quad \text{and} \quad z_1 = m~,~~ z_2=-m~,~~ z_3 = -\frac{1}{2\alpha}~.
\end{equation}
The coordinate transformations between the spherical-like and Weyl coordinates are \cite{griffiths_podolsky_2009}:
\begin{equation}
\rho = \frac{r\sin\theta ~\sqrt{PQ}}{\left(1-\alpha r\cos\theta\right)^2} \qquad \text{and} \qquad z = \frac{\left(\cos\theta-\alpha r\right)\left[r-m\left(1+\alpha r\cos\theta\right)\right]}{\left(1-\alpha r \cos\theta\right)^2}~.
\end{equation}
In the Newtonian picture that one can associate with Weyl solutions (not to be confused with the Newtonian limit; see Section \ref{linesource}), the C-metric can be seen as ``sourced" by the Newtonian potential --- the function $u(\rho,z)$ in equation \eqref{weyl-u} --- of the superposition of a finite line mass and a semi-infinite line mass (see Figure \ref{Newtonian}), as was first shown by Bonnor \cite{Bonnor}. 

Starting with the static region $(2m<r<1/\alpha)$, and by introducing null coordinates, Kinnersley and Walker extended the spacetime through both the horizons to include the regions $0< r \leq 2m$ and $r\geq 1/\alpha$, and, more importantly, to include the causally disconnected second black hole \cite{Kinnersley_Walker}. When extending the spacetime, one has some freedom of choosing the topology. The conventional choice is to choose the simply connected topology. However, one can instead choose a nontrivial topology by identifying a pair of static regions in which case the two black hole horizons of the C-metric can be interpreted as the opposite ``mouths'' of a wormhole (see section 6 of \cite{Kinnersley_Walker}). They also construct 2D conformal diagrams of the extended spacetime, however, they do not construct $\mathscr{I}$ explicitly. Ashtekar and Dray subsequently showed that the C-metric is asymptotically flat at null infinity $\mathscr{I}$ by explicitly constructing $\mathscr{I}$ and showing that it has the topology $\mathbb{R}\times \mathrm{S}^2$ (see Section \ref{conformal_diagrams}) \cite{Ashtekar_Dray}. They further showed that the C-metric has nonzero Bondi news --- the first exact solution to have so.
Griffiths et al.~{ \cite{griffiths_krtous_podolsky}} used the simpler root structure of Hong-Teo coordinates \cite{Hong_Teo} to analyze the C-metric in an algebraically simpler fashion. They also used Hong-Teo coordinates to construct Kruskal-Szekeres coordinates and to construct 2D and 3D conformal diagrams. The two black holes can also be seen simultaneously using the boost-rotation symmetric form of the metric (see Section \ref{bs-section}), although the interiors of the horizons are not covered in these coordinates. We shall be using the above coordinate systems, that is, \eqref{C-metric-HongTeo}--\eqref{C-metric-HongTeo-func} and \eqref{c-metric-metric}--\eqref{c-metric-func} interchangeably, along with the Weyl and boost-rotation symmetric coordinates.

\subsection{Conical singularities and line sources}\label{con-sing-sec}
Given a 2-dimensional manifold, local flatness around a point would imply that the ratio of circumference to radius of small circles around the point must be $2\pi$, the failure of which is referred to as a conical singularity (or a nodal singularity).
These ratios for the C-metric at $\theta = 0$ and $\theta = \pi$ are $2\pi C (1-2\alpha m)$ and $2\pi C (1+2\alpha m)$, respectively,
implying conical defects on the axis. The deficit angles ($\delta_i \equiv 2\pi - \vartheta_i$ where $\vartheta_i$ are the total angles on the axis) are $2\pi - 2\pi C(1-2\alpha m)$ and $2\pi - 2\pi C (1+2\alpha m)$. For any value of $C$, both singularities cannot be simultaneously removed. It is customary to choose $C = 1/(1+2\alpha m)$ so that the negative $z$-axis ($\theta = \pi$) is regular. This choice is inspired by the embeddability of the horizons (see Section \ref{embedding_section}). The deficit angles in this case are $\delta_0 = \frac{8\pi \alpha m}{1 + 2\alpha m} $ and $\delta_\pi = 0$.

\begin{figure}[htb]
\begin{subfigure}[t]{0.3\textwidth}
    \centering
    \includegraphics[height=2cm]{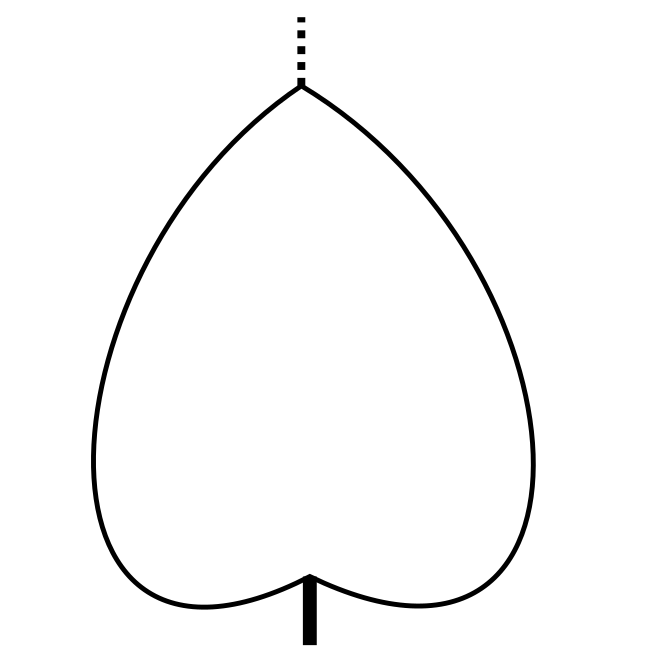}
    \caption{Surface with string and strut, $C=1$.}
    \label{fig:strut_and_string}
\end{subfigure}\hfill
\begin{subfigure}[t]{0.3\textwidth}
    \centering
    \includegraphics[height=2cm]{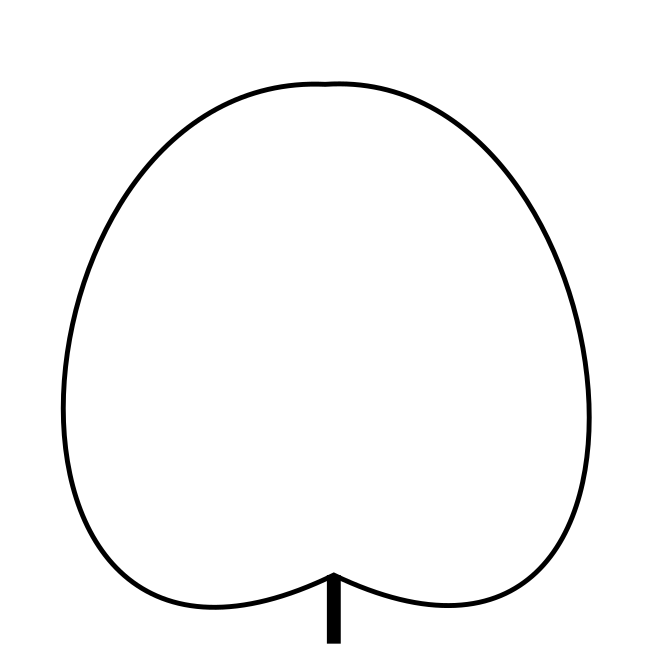}
    \caption{Choosing $C$ such that top axis is regular, ${C^{-1}={1-2\alpha m} }$. Surface with only strut.}
    \label{fig:strut}
\end{subfigure}\hfill
\begin{subfigure}[t]{0.3\textwidth}
    \centering
    \includegraphics[height=2cm]{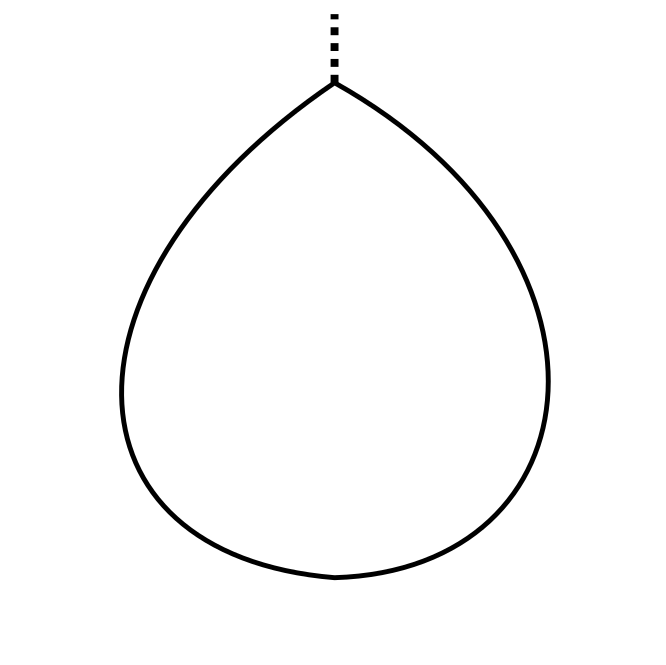}
    \caption{Choosing $C$ such that bottom axis is regular, ${C^{-1}={1+2\alpha m} }$. Surface with only string.}
    \label{fig:string}
\end{subfigure}
\caption{Representation $ t,r=\text{constant}$ surfaces with various choices for the range parameter $C$. Dotted lines represent strings and thick lines represent struts.}
\label{conic_figure}
\end{figure}

It is well known that deficit angles represent a string ``pulling'' on the horizon and that excess angles represent a strut ``pushing'' on the horizon \cite{Hong_Teo, griffiths_podolsky_2009}. The ``natural'' choice of C-metric with parameter $C=1$ (i.e.~$\varphi$ has period $2\pi$) will have both deficit and excess angles, meaning it has both a string and a strut. By choosing $C=\frac{1}{1+2\alpha m}$, the strut can be eliminated, but at the cost of increasing the strength of the string. Similarly, by choosing $C=\frac{1}{1-2\alpha m}$, the string can be eliminated, but at the cost of increasing the strength of the strut. Figure \ref{conic_figure} shows the strings and struts of the C-metric for different choices of range $C$.

For a Weyl metric of the form \eqref{weyl-metric}, Israel gave the energy-momentum density of the line source on the axis \cite{Israel}
\begin{equation}\label{eq:linesource}
\mathcal{L}_2^2 = \mathcal{L}_4^4 = \frac{1}{4}\left( 1 - e^{\nu_0} \right)~,
\end{equation}
where $\nu_0 \equiv \nu(\rho=0,z)$. However, the parameter $C$ hidden in the coordinate range of $\varphi\in [0,2\pi C)$ will alter the formula for the line source. Rescaling $\varphi \rightarrow \frac{\varphi}{C}$ will make the parameter $C$ appear explicitly in the metric. To maintain the Weyl form under this rescaling, it is required to change $u\to u +\ln(C)$, $\nu \rightarrow \nu + \ln(C)$, and $t\to \frac{t}{C}$.
Hence, the new functions are given by
\begin{gather}
e^{2u} = \frac{1}{C^2} \alpha \frac{R_1+R_2-2m}{R_1+R_2+2m}\left( R_3 + z + \frac{1}{2\alpha} \right)~, \label{eq:modified-u} \\[5pt]
e^{2\nu} = \frac{1}{C^2}\alpha \frac{R_1+R_2-2m}{R_1+R_2+2m}\left( R_3 + z + \frac{1}{2\alpha} \right) \frac{\bigl[ (1-2\alpha m)R_1 + (1+2\alpha m)R_2 + 4\alpha m R_3 \bigr]^2}{8\alpha (1-4\alpha^2m^2)^2R_1R_2R_3}~. \label{eq:modified-nu}
\end{gather}
This gives the line source
\begin{equation}
\mathcal{L}^2_2 = \mathcal{L}^4_4 = \quad \begin{cases} \frac{1}{4} \left(1-\sqrt{\frac{1 - 2\alpha m}{1 + 2\alpha m}}\right) &\text{if}~~ z>m\\[6pt] \frac{1}{4} \left(1-\sqrt{\frac{1 + 2\alpha m}{1 - 2\alpha m}}\right) &\text{if}~~ -\frac{1}{2\alpha}<z<-m \end{cases}~.
\end{equation}

\subsection{The Gauss-Bonnet theorem} \label{gb} 
For a 2-dimensional manifold $M$ in the presence of a boundary and/or conical singularities, the (modified) Gauss-Bonnet (GB) formula is \cite{EGUCHI}
\begin{equation}\label{eqn:gauss-bonnet}
    2\pi\chi(M) = \int_M K dA + \int_{\partial M} k_g ds - \sum_{i} \beta_i
\end{equation}
where $\chi(M)$ is the Euler characteristic -- an integer-valued topological invariant, $K$ is the Gaussian curvature, $k_g$ is the geodesic curvature on the boundary and $\beta_i$ are the excess angles measured at the conical singularities ($\beta_i \equiv \vartheta_i - 2\pi$ where $\vartheta_i$ are the total angles on the axis).

This formula can be used to detect the presence of a boundary or the presence of conical singularities. Since the Euler characteristic must be an integer, if the total curvature $(\equiv \int_M K dA)$ evaluates to a non-integer, then there must be a boundary or conical singularities (or both) present. However, the converse is not necessarily true, as we shall see below.

The $r$-constant, $t$-constant surfaces of \eqref{c-metric-metric} -- \eqref{c-metric-func} (for $r<1/\alpha$) are given by\footnote{The conformal factor can be ignored as it is smooth and regular for $\theta\in [0,\pi]$. Note that the Gaussian curvature will change if the conformal factor is ignored, but the total curvature does not, due to the opposite change in the area element $dA$.}
\begin{equation}
ds^2 = \frac{1}{P}d\theta^2 + P \sin^2\theta d\varphi^2 ~,
\end{equation}
and the Gaussian curvature for such surfaces is given by
\begin{equation}\label{c-gaussian}
K_0 = 1-2\alpha m\cos\theta~.
\end{equation}
The total curvature is
\begin{equation}
\int_{M} K_0 dA = \iint K_0 \sin\theta  d\theta d\varphi = 4\pi C~,
\end{equation}
which is not an integer multiple of $2\pi$, showing the presence of conical singularities (and potentially a boundary). However, since the sum of excess angles is $4\pi C - 4\pi$, the modified GB theorem\footnote{ For $r<1/\alpha$, there is no boundary term since $\theta$ takes its full range. The case $r\geq 1/\alpha$ is discussed in Section \ref{surface-geometry}.} gives $\chi=2$. Thus, the $r$-constant, $t$-constant surfaces ($r<1/\alpha$) are two-spheres even in the presence of conical singularities. 
In general, in the presence of conic angles (and no boundaries), in a 2-metric of the form
\begin{equation}
ds^2 = f(\theta) d\theta^2 + \sin^2\theta d\phi^2~,
\end{equation}
where $\phi\in[0,2\pi C)$,
the condition for it to be topologically $\mathrm{S}^2$ works out to be (see Appendix \ref{top_s2})
$ C\left[ \sqrt{f(0)^{-1}} + \sqrt{f(\pi)^{-1}} \right] = 2$. For the C-metric (equation \eqref{c-metric-metric}), $f = 1/CP$, hence, 
\begin{equation}
\frac{1}{\sqrt{f}|_0} + \frac{1}{\sqrt{f}|_\pi} = \frac{1-2\alpha m}{C} + \frac{1+2\alpha m}{C} = \frac{2}{C}~,
\end{equation}
satisfying the above condition, as expected.

\subsubsection*{The Converse}
The C-metric presents us with a curious example where the total curvature $\int_M K dA$ can be $2\pi \times\mathrm{integer}$ even in the presence of conical singularities. The sum of the excess angles is $4\pi C - 4\pi$ and hence, vanishes in the case $C=1$.   
In other words, there is an excess angle and a deficit angle of equal magnitude.

Often the converse of the original Gauss-Bonnet theorem
\begin{equation}\label{eqn:gauss-bonnetoriginal}
    2\pi\chi(M) = \int_M K dA 
\end{equation}
is used for regular two-spheres. For example, this has been used to obtain regular local black holes that have two-sphere horizon topology \cite{LocalToroidalXanthapolous}. The C-metric shows how this can go wrong. If one innocuously applies (\ref{eqn:gauss-bonnetoriginal}), one would pick up the $C=1$ C-metric as a regular local black hole.

\subsection{Bondi Coordinates}
Bondi-Sachs coordinates, introduced in \cite{bondi1962, sachs1962_waves}, are naturally adapted to describing the metric structure at null infinity of asymptotically flat spacetimes. To get a physical picture of these coordinates, imagine a light source in the spacetime. The coordinates $\mathfrak{u}, r, \theta, \phi$ are chosen such that only the coordiante $r$ varies along a light ray. The surfaces described by $\mathfrak{u}$-constant $r$-constant are the wavefronts of the light. The coordinates $\theta$ and $\phi$ are chosen such that they define a $\mathrm{S}^2$ submanifold with the desired determinant, that is, $r^4\sin^2\theta$. The general form of the metric for an asymptotically flat spacetime in these coordinates is
\begin{equation}\label{Bondigeneral}
    ds^2 = \left(\frac{V}{r} e^{2\beta} - r^2 U^2 e^{2\gamma}\right) d\mathfrak{u}^2 + 2e^{2\beta} d\mathfrak{u} dr + 2r^2 U e^{2\gamma} d\mathfrak{u} d\theta -r^2 (e^{2\gamma }d\theta^2  + e^{-2\gamma} \sin^2\theta d\phi^2)
\end{equation}
where the functions $U, V, \beta,$ and $\gamma$ satisfy the fall-off conditions
\begin{equation}\label{eq:fall-off-conditions}
\lim_{r\to \infty} U = \lim_{r\to \infty} \beta = \lim_{r\to \infty} \gamma = 0 \quad \text{and} \quad \lim_{r\to\infty} \frac{V}{r} = 1~,
\end{equation}
but are otherwise arbitrary functions of the coordinates.

Bonnor \cite{Bonnor_1990} made the curious observation that with the substitution $x=\cos\theta$ the null form of the C-metric (\ref{null-coordinates}) acquires the Bondi-Sachs form
\begin{equation}
\label{c-metric-null-theta}
    ds^2 = H d\mathfrak{u}^2 + 2d\mathfrak{u} dr - 2\alpha r^2 \sin\theta d\mathfrak{u} d\theta - r^2(B^{-1} \sin^2\theta d\theta^2 + Bd\phi^2),
\end{equation} with: 
\begin{align}
    V&= -\alpha r^2 (2\cos\theta + 6\alpha m \cos^2\theta) + r(1+6\alpha m \cos\theta) -2m\\
    e^{2\gamma} &= \frac{1}{1-2\alpha m \cot^2\theta \cos\theta} \\
    U &= -\alpha (1-2\alpha m \cot^2\theta \cos\theta)\\
    e^{2\beta} &=1.
\end{align}
However, although he achieved the Bondi form, these functions do not satisfy the fall-off conditions \eqref{eq:fall-off-conditions}, especially $V$ has positive powers in $r$. Bonnor ascribes this to the acceleration of the C-metric. 

\subsubsection*{Standard Bondi coordinates for the C-metric (through the Weyl form)}
Bonnor further comments that the ``standard" Bondi form in which the functions have correct fall-off conditions can be obtained since the C-metric is asymptotically flat and also has a Weyl form, referring to the procedure outlined in the appendix of Bondi et al. \cite{bondi1962}, and commented that such a form ``must be very complicated." We now show that finding standard Bondi coordinates for the C-metric is not possible through the Weyl form, which we believe was never tried before.

The procedure, as outlined in Bondi et al.,  begins with a Weyl metric (such as \eqref{weyl-metric}) and uses the transformations $\rho = R\sin\Theta$ and $z=R\cos\Theta$, after which the metric takes the form
\begin{equation}
    ds^2 =   -e^{2u}dt^2 + e^{2\nu-2u}\left( dR^2 + R^2d\Theta^2 \right) + R^2 \sin^2\Theta e^{-2u} d\phi^2~.
\end{equation}
Then by applying the transformations,
\begin{equation}
    t = \mathfrak{u} + f(R,\theta)~, \qquad \Theta = \Theta(R,\theta)~,
\end{equation}
and by enforcing $g_{11} = g_{12} = 0$ (cf. equation \eqref{Bondigeneral}) and then eliminating $f$, they find
\begin{equation}
\label{eqn:Theta}
    R^2\Theta_{,R} \left( e^{2\nu-4u} \right)_{,\theta} = \Theta_{,\theta} \left(e^{2\nu-4u} \frac{R^4\Theta^2_{,R}}{1+R^2\Theta^2_{,R}}\right)_{,R}~
\end{equation}
where the subscripts denote partial derivatives. They then require the following fall-off condition for asymptotic flatness
\begin{equation}\label{eq:Weyl-expansion}
    e^{2\nu -4u} = 1 + \frac{4m}{R} + \frac{p(\Theta)}{R^2} + \frac{q(\Theta)}{R^3} + \cdots~
\end{equation}
and the condition
\begin{equation}\label{eq:Weyl-static}
    \lim_{R\rightarrow \infty} \Theta_{,\theta} =1.
\end{equation}
{  The condition \eqref{eq:Weyl-expansion} is inspired from the behavior of known asymptotically flat Weyl solutions such as the Schwarzschild solution. 
It is easy to check, for example, that for the Schwarzschild metric in Weyl form (see, for instance, \cite{stephani_exact_sol}) it has the following expansion:
\begin{equation}\label{eq:Weyl-Schwarzschild}
e^{2\nu-4u} = 1 + \frac{4m}{R} + \frac{m^2(\cos^2\Theta + 7)}{R^{2}} + \frac{6m^3 (\cos^2\Theta + 1)}{R^{3}} + \mathcal{O}\left(\frac{1}{R^4}\right)~,
\end{equation}
which gives $p(\Theta) = m^2(\cos^2\Theta + 7)$, $q(\Theta) = 6m^3 (\cos^2\Theta + 1)$, etc.
The condition \eqref{eq:Weyl-static} ensures that there is no radiation in the spacetime (since we are dealing with static spacetimes). One expands the functions $f$ and $\Theta$ in powers of $R$, and solves for $R$ by the condition that the area of the spherical part remains invariant: $r^4\sin^2\theta = g_{22} g_{33}$.} Then, to satisfy (\ref{eqn:Theta}), $\Theta$, $f$, and $R$ should have the following expansions:
\begin{align}
    \Theta &= \theta + \frac{p'}{4R^2} + \frac{q'-6mp'}{12R^3} + \cdots \\
    f &= R + 2m\log R - \frac{\frac{1}{2}p-2m^2}{R} - \frac{\frac{1}{2}q-mp+4m^3}{2R^2} + \cdots \\
    R &= r-m-\frac{1}{2} \left( \frac{1}{4}p'\cot\theta+\frac{1}{2}p-3m^2+\frac{1}{4}p'' \right) r^{-1}  \\ 
    &\hspace{2em} -\frac{1}{2} \left( \frac{1}{12}q''+\frac{1}{12}q'\cot\theta+\frac{1}{2}q-mp+4m^3 \right)r^{-2}+\cdots \nonumber .
\end{align}
However, for the C-metric, the expansion of the metric function $e^{2\nu-4u}$ takes the form
\begin{equation}
\label{eqn:expansion-c-metric}
{ e^{2\nu-4u}} = \frac{1}{{ \left(1+\cos\Theta\right)}2 \alpha^{2} \eta^{2} R^{2}} + \frac{4 \alpha  m-1}{4 \alpha^{3} \eta^{2} R^{3}} + \frac{3\cos\Theta(1-4\alpha m \eta) - 4\alpha m \eta}{16 \alpha^4 \eta^{2} R^4 } + \mathcal{O}\left( \frac{1}{R^5} \right)~,
\end{equation}
where $\eta = 1 - 2\alpha m$, $0<2\alpha m <1$, and $m>0$. Since this expansion starts at the second order, Bondi's formula for $\Theta$ will not apply. Thus, one has to find a solution for $\Theta$, by substituting the expansion from equation \eqref{eqn:expansion-c-metric} in equation \eqref{eqn:Theta} and solve for the expansion coefficients. However, we find that this gives the result $\Theta=\theta$ identically, with all higher-order terms having coefficient zero. This clearly contradicts the defining property for $\Theta$, ${ g_{11}= {g_{12}}=0 }$. Thus, one cannot find a standard Bondi form of the C-metric through the Weyl form following Bondi's prescription.

To summarize, although the C-metric admits Weyl coordinates in the static region and is asymptotically flat, the fact that the static region does not extend to infinity implies that the Weyl form of the C-metric cannot be used to find standard Bondi coordinates. This is where Bonnor's expectation was mistaken. (As we shall discuss later, Figure \ref{fig:informal-diagram-2} shows the Weyl coordinate curves, which clearly illustrate that the limit $\rho\to\infty$ is just one particular point on $\mathscr{I}$, and does not cover the entire null infinity.)

However, a standard Bondi-Sachs form for the C-metric does exist, as shown in \cite{Sladek-Finley}. The Bondi mass aspect has been explicitly computed in \cite{Sladek-Finley} following a general procedure outlined in \cite{Tafel-Pukas, Tafel, Natorf-Tafel} in which one starts with the standard null coordinate \eqref{c-metric-null}, and sequentially chooses appropriate conformal factors.

\section{The Generalized C-metric}\label{ernst-c-metric}
Ernst's solution-generating technique generates a one-parameter extension of any vacuum spacetime with two commuting hypersurface orthogonal Killing fields \cite{Ernst1978, Ernst1979}. As mentioned in the introduction, Ernst introduced it to regularize the C-metric. He conjectured that for a particular value of the external field parameter $k$, the spacetime will be free from conical singularities.

Given a static axisymmetric vacuum solution of Einstein's equations of the form
\begin{equation}\label{metric1}
ds^2 = - fdt^2 + h(dx_1^2+dx_2^2) + ldx_3^2
\end{equation}
where $f$, $l$ and $h$ are functions only of $x_1$ and $x_2$, $\partial_t$ and $\partial_{x_3}$ are the two commuting, hypersurface orthogonal Killing vector fields, Ernst's prescription gives a new family of vacuum solutions 
\begin{equation}
ds^2 = - fe^{k(F+L)} dt^2 + he^{k(F-L)-k^2lf}(dx_1^2+dx_2^2) + le^{-k(F+L)}dx_3^2
\end{equation}
where $k\in (-\infty, \infty)$, and $L$, $F$ are real auxiliary functions solving
\begin{equation}\label{LF}
\nabla L = \mathrm{i}\sqrt{fl} \nabla(\ln l)~, \qquad \nabla F = \mathrm{i}\sqrt{fl} \nabla(\ln f)~,
\end{equation}
and $\nabla = \frac{\partial}{\partial x_1} + \mathrm{i}\frac{\partial}{\partial x_2}.$ Applying this to the polar form of the C-metric \eqref{c-metric-metric}--\eqref{c-metric-func}, we obtain
\begin{equation}
ds^2 = \frac{1}{(1-\alpha r \cos\theta)^2}\left[ - Q e^{k(F+L)}dt^2 + e^{k(F-L)- k^2 W} \left( \frac{dr^2}{Q} + \frac{r^2 d\theta^2}{P}\right)  + {P r^2 \sin^2\theta} e^{-k(F+L)}d\varphi^2 \right]
\end{equation}
where
\begin{equation}
P = 1-2\alpha m \cos\theta~, \qquad Q = \left(1-\frac{2m}{r}\right)\left(1-\alpha^2 r^2\right)~,
\end{equation}
\begin{subequations}\label{FL}
\begin{equation}
F(r,\theta) = -\frac{\alpha P r^2 \sin^2\theta}{(1-\alpha r \cos\theta)^2} + 2m\cos\theta~, \qquad L(r,\theta) =  \frac{Q}{\alpha(1-\alpha r \cos\theta)^2} + \frac{2m}{\alpha r}~,
\end{equation}
and 
\begin{equation}
W(r,\theta) = \frac{P Q r^2 \sin^2\theta }{(1-\alpha r \cos\theta)^4}~.
\end{equation}
\end{subequations}
If we apply Ernst's prescription to \eqref{C-metric-HongTeo} instead, there will be a factor of $\alpha$ difference: $F' = F/\alpha$, $L'= L/\alpha$ and $W'= W/\alpha^2$. This is simply because $\tau=\alpha t$ (see equation \eqref{HT-sph-trans}) which changes the right hand side of \eqref{LF} by this factor. There is no practical difference because this can be absorbed in $k$. 
\subsubsection*{Petrov Type}
By computing the Weyl scalar invariants $I$ and $J$ \cite{stephani_exact_sol}, one can check that the generalized C-metric is of Petrov type I, meaning that the algebraic speciality of the C-metric is lost for all values of $k\ne 0$.

\subsubsection*{Removing Conical Singularities: The Regularized C-metric}
The ratios of circumference to radius of small loops around the axis for $\theta = 0$ and $\theta= \pi$ are given by $2\pi C (1-2\alpha m) e^{ - 2mk }$ and $2\pi C (1+2\alpha m) e^{ 2mk }$, respectively.
To remove both singularities, the expressions in the above two equations must be equal to $2\pi$, requiring
\begin{equation}\label{c-sing}
(1-2\alpha m) e^{- 2mk} = (1+2\alpha m) e^{ 2mk} ~=~ \frac{1}{C}~.
\end{equation}
This has a unique solution:
\begin{equation}\label{eqn:k-value}
k_0 = \frac{1}{4m} \ln \frac{1-2\alpha m}{1 + 2\alpha m}~, \qquad C = \frac{1}{\sqrt{1-4\alpha^2 m^2}}~.
\end{equation}
Thus the conical singularities of the new metric can be simultaneously removed. We want to emphasize that having the correct value of $k$ and $C$ is very important. Our values differ from those of Dray and Walker \cite{TDray_MWalker} and Hogan \cite{Hogan_1979}, but are in agreement with Bonnor (who used Weyl coordinates) \cite{Bonnor88}. In Appendix \ref{Dray-Walker-connection} we show that Dray and Walker's value, obtained in Kinnersley-Walker coordinates, can be obtained from ours by rescaling and coordinate transformation, after fixing an error in their result.

\subsubsection*{Weak Field Limit}
It is easy to verify that applying Ernst's prescription to the C-metric and then setting $m=0$ gives the same result as setting $m=0$ first and then applying Ernst's prescription. For $m=0$, the C-metric remains conically singular unless one chooses $C=1$ --- this is Rindler spacetime. Applying Ernst's prescription will not change its regularity for any value of $k$. For the regularized C-metric, note that as $ m \to 0$ the parameter $k$ in equation \eqref{eqn:k-value} approaches $ -\alpha$, and $C$ approaches $1$, giving $\varphi $ a periodicity of $2\pi$. Thus, in the zero-mass limit, the regularized C-metric returns a particular generalization of the Rindler metric.

\subsection{Curvature Scalars}
Similar to the original C-metric, the generalized C-metric has $w_1$ and $w_2$ as the only nonvanishing curvature scalars (see equation \eqref{eq:curvaturescalars}). The corresponding expressions for the scalars $w_1$ and $w_2$ are very large, and the relevant Maple code is given in Appendix \ref{sec:w1w2}. However, after analyzing them, one finds that in addition to the curvature singularity at $r=0$ (i.e., $y=\infty)$, there are three additional singularities at $y= -1, 1$ and $1/2\alpha m$ with $x=-y$. The last one at $y=1/2\alpha m$, is outside the physical spacetime.
The new singularities occur since Ernst's prescription introduces a dipole gravitational field (Newtonian potential $kz$ in Weyl coordinates, see below) at infinity while resolving the conical singularities.

\begin{figure}[htb]
\centering
\includegraphics[width=0.5\textwidth]{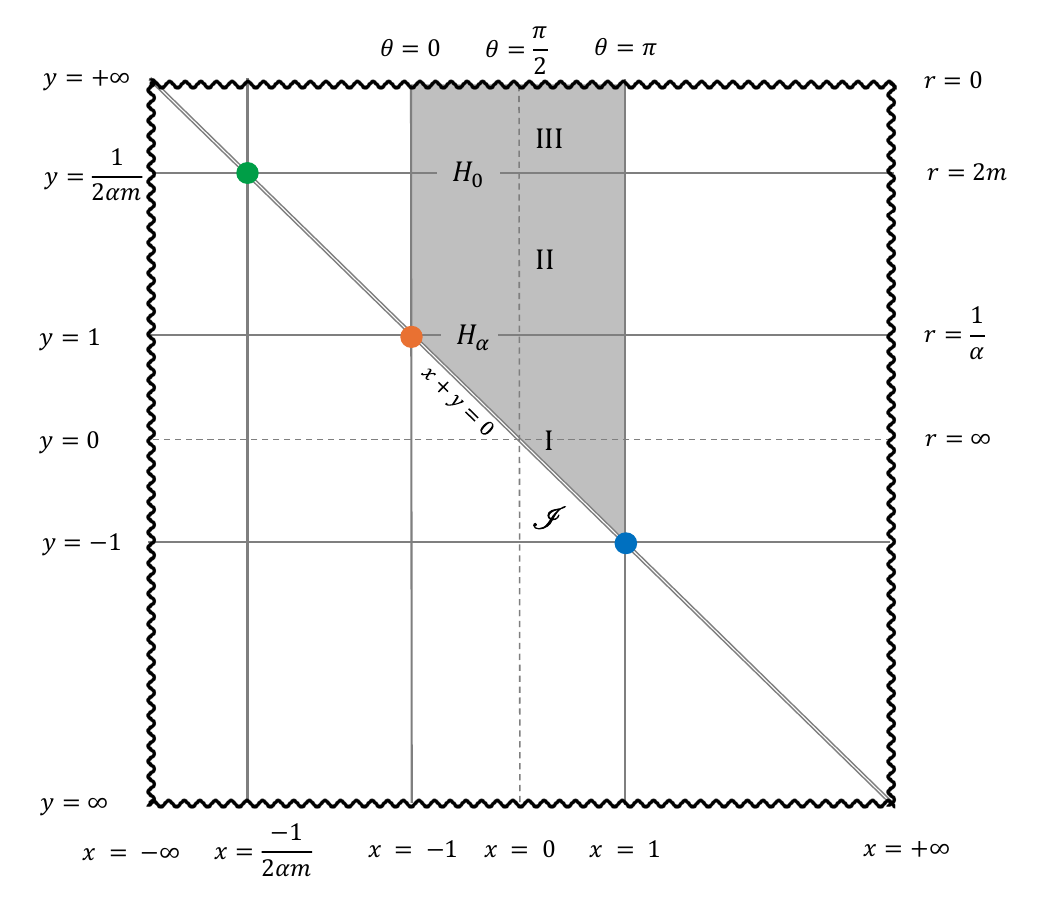}
\caption{Coordinate diagram representing the three new singularities (large colored dots) of the distorted C-metric. These points are regular for the C-metric. Note that the singularity at $(x,y) = (-1,1)$ is $(r,\theta)=(1/\alpha, 0)$ and the one at $(x,y)=(1,-1)$ is not covered by the $r$ coordinate. The singularity at $y=1/2\alpha m$ is outside the physical spacetime.}
\label{CoordPlot}
\end{figure}

\subsubsection*{Type of sources at infinity}
Like the original C-metric, the generalized C-metric can be seen as a distorted/local black hole \cite{GerochHartle, chandrasekharbook, LocalToroidalXanthapolous}. In Weyl coordinates, the general form of a distorted black hole is given by \cite{Astorino21}
\begin{equation}
    ds^2 = e^{2u_0 + 2u_{ext}} dt^2 + e^{2(\nu_0 + \nu_{ext}) - 2(u_0 + u_{ext})} \left( d\rho^2 + dz^2 \right) + \rho^2 e^{-2u_0 - 2u_{ext}} d\phi^2~,
\end{equation}
where, $u_0$ and $u_{ext}$ are the Newtonian potentials (see Section \ref{new.inter.}) of the background black hole and the external gravitational fields, respectively, and $\nu_0$ is obtained through quadrature by solving the differential equations
\begin{equation}\label{nu-equation}
\nu_{0,\rho} = \rho\left( u_{0,\rho}^2 - u_{0,z}^2 \right) \qquad \text{and} \qquad \nu_{0,z} = 2 u_{0,\rho} u_{0,z}.
\end{equation}
The field $u_{ext}$ is given by
\begin{equation}
    u_{ext} = \sum_{n=1}^\infty a_n \mathcal{R}^n P_n
\end{equation}
where $a_n$ are real constants that are related to the multipole momenta of the external fields \cite{Astorino21}, ${\mathcal{R} = \sqrt{\rho^2 + z^2}}$, and $P_n \equiv P_n\left(\frac{z}{\mathcal{R}}\right)$ are Legendre polynomials, and $\nu_{ext}$ can be obtained by solving the same differential equations in \eqref{nu-equation}, with $\nu_0+\nu_{ext}$ replacing $\nu_0$, and $u_0 + u_{ext}$ replacing $u_0$. The generalized C-metric is obtained by choosing $a_1=k$ and all other constants zero. Thus, the new singularities of the generalized C-metric can be thought of as dipole sources at infinity. These singularities affect the conformal structure of the regularized C-metric, as we will explore in detail in Section \ref{asym_str}. In particular, the topology of the null infinity $\mathscr{I}$ will no longer be $\mathbb{R}\times \mathrm{S}^2$, and the spacetime is no longer asymptotically flat. Despite this, as we will see, one can extend the spacetime to include the causally disconnected second black hole in a consistent way, just as in the case of the original C-metric. This means that it may be possible to create semiclassical black hole pairs with appropriate points excised from the Euclideanized spacetime, as in the case of other C-metrics; however, this falls beyond the scope of the present work.

\subsection{Surface geometry}\label{surface-geometry}
The Gaussian curvature for the $r$-constant, $t$-constant surfaces of the generalized C-metric is
\begin{equation*}
\begin{split}
&K = K_0 + \frac{k}{\alpha ^2 (x+y)^4} \left[B \left(-2 \alpha  m (5 x+14 y)+5 x y+17\right)+3 \left(4 \alpha ^2 m^2 (5 x^2+4 x y-1) -14 \alpha my (x^2-1) \right.\right. \\ & \left.\left. +  x^3y-5xy +3 x^2 -7\right)\right] + \frac{4 k^2 B}{\alpha ^4 (x+y)^6} \left[-4 \alpha ^2 m^2 \left(x^2+2 x y+1\right)+4 \alpha  m (x^2 y+2x+y)- x^2 +2xy-1\right]~,
\end{split}
\end{equation*}
where $K_0$ is the Gaussian curvature for the C-metric, given in equation \eqref{c-gaussian}. 

{  Notice that these surfaces do not have a boundary ($\theta$ takes the full range of values) and do not have conical singularities for $k=k_0$. Using this curvature expression to evaluate the integral in the Gauss-Bonnet theorem precisely gives (using Mathematica)
\begin{equation}
\int K dA = 2 \pi C \left[ (1+2\alpha m)e^{2k_0m} + (1-2\alpha m)e^{-2k_0m} \right] = 4\pi~.
\end{equation}
This holds for any $r<1/\alpha$ and hence, Gauss-Bonnet implies the surfaces are two-spheres. For  $ k\neq k_0$, the metric will have conical singularities, and contributions from the conical angles have to be considered to show that these surfaces are still two-spheres as in the case of the original C-metric, Section \ref{con-sing-sec}. }

\begin{figure}[htb]
    \centering
    \includegraphics[width=\textwidth]{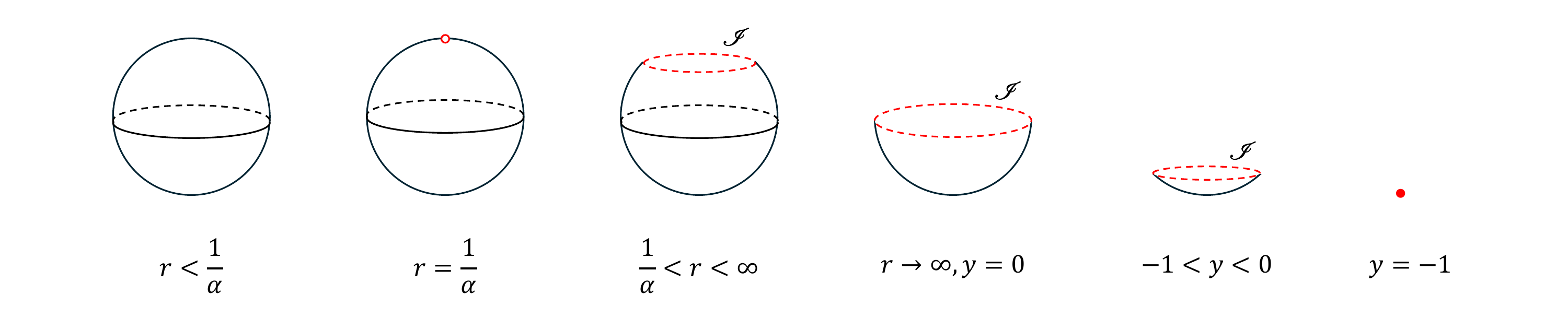}
    \caption{Representation of $r$-constant, $t$-constant surfaces, for $r\geq 1/\alpha$ for the generalized C-metric (same for the C-metric). The red lines (or dots) represent the boundary of the surfaces.}
    \label{fig:half-moon}
\end{figure}

\begin{figure}[htb]
    \begin{subfigure}[b]{0.49\textwidth}
    \centering
    \includegraphics[width=0.82\textwidth]{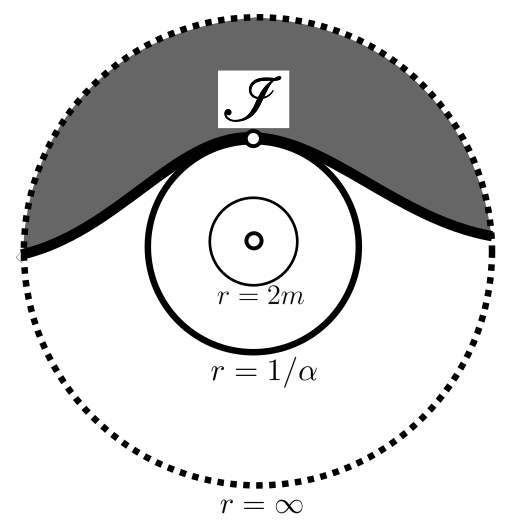}
    \caption{``Informal diagram" showing the location of $\mathscr{I}$ for different values of $\theta$.}
    \label{fig:informal-diagram}
    \end{subfigure}\hfill
    \begin{subfigure}[b]{0.49\textwidth}
    \centering
    \includegraphics[width=\textwidth]{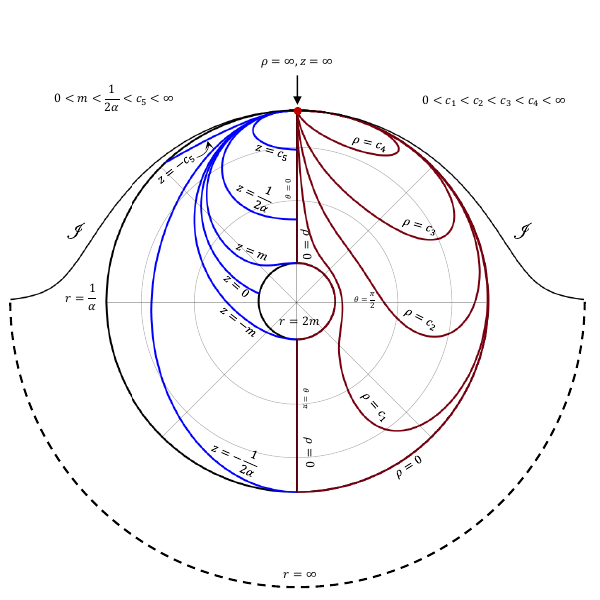}
    \caption{Diagram showing the $\rho$-constant and $z$-constant Weyl coordinate curves in the informal diagram.}
    \label{fig:informal-diagram-2}
    \end{subfigure}
    \caption{Figure \ref{fig:informal-diagram} is what we call the ``informal diagram''  of the regularized C-metric as covered by the $r$ coordinate. It shows the location of $\mathscr{I}$ for different values of $\theta$ in terms of $r$. The shaded part of the figure is not a part of the spacetime.  Note that for $\theta>\frac{\pi}{2}$, $\mathscr{I}$ cannot be reached by any value of the $r$ coordinate. Equivalently, for $r>1/\alpha$, the range of $\theta$ is limited by $\mathscr{I}$ as shown by the semicircle which corresponds to a partial sphere in Figure \ref{fig:half-moon}. The curvature singularity at the point $(r,\theta) = (1/\alpha, 0)$ is shown as a hole on $\mathscr{I}$. The central hole is the $r=0$ singularity. This diagram is the same for C-metric except that $(r,\theta) = (1/\alpha, 0)$ is not a curvature singularity. Figure \ref{fig:informal-diagram-2} shows different Weyl coordinate curves on the informal diagram. The $\rho$-constant curves are shown in red on the right half of the figure, and $z$-constant curves are shown in blue on the left half of the figure. These curves are plotted on a $\varphi$-constant slice, and the full 3D surface can be obtained  by rotating the curves about the axis. Both $\rho\to\infty$ and $z\to\infty$ curves reduce to the point $(r,\theta) = (1/\alpha, \pi)$. }
\end{figure}

For $r\geq 1/\alpha$ the metric diverges at $r = \frac{1}{\cos\theta}$ which limits the permissible coordinate range for $\theta$ in the integral and introduces a boundary. In this case, the surface will have topology $\mathbb{R}^2$ due to the introduction of a boundary. Figure \ref{fig:half-moon} shows the boundary for different values of $r$. Since $\chi(\mathbb{R}^2)=1$, the surface topology for $r>1/\alpha$ could be explicitly verified by again using the Gauss-Bonnet theorem with the boundary term  giving $2\pi$:
\begin{equation}
    \int_M KdA + \int_{\partial M} k_g ds = 2\pi
\end{equation}
where, for $\theta=\text{constant}$, the geodesic curvature is given by $\int k_g ds = -\pi C \frac{g_{\theta \theta, \theta}}{\sqrt{g_{\theta \theta} g_{\phi \phi}}}.$

The expression for the surface area of a general $t$-constant $r$-constant surfaces is given by
\begin{equation}
A = \frac{2 \pi  r^2}{\sqrt{1-4 \alpha ^2 m^2}} \int_0^{\pi} \frac{\sin\theta ~e^{\left(-\frac{1}{2} k_0^2 W - k_0 L\right)}}{(1-\alpha  r \cos (\theta ))^2} d\theta~.
\end{equation}
This converges for $r<1/\alpha$. For $r\geq1/\alpha$, the integrand diverges at ${ \cos\theta=\frac{1}{\alpha r} }$, and it is easily checked that the area integral diverges as well. As we will see in Section \ref{asym_str}, this happens because $r=\frac{1}{\alpha \cos\theta}$ is $\mathscr{I}$, the conformal infinity. Figure \ref{fig:surface-area} compares the surface areas of the generalized C-metric and the (original) C-metric.
The area of the black hole horizon ($r=2m$) is given by
\begin{equation}
A  =  e^{-\frac{k}{\alpha}} \frac{16\pi m^2 C}{1-4\alpha^2 m^2}~.
\end{equation}

\subsubsection*{Circumferences}
Integrating the line element for $t, r=\text{const.}$ and $\theta=\pi/2$,  the equatorial circumference works out to be
\begin{equation}
    \mathcal{C}_{\mathrm{eq}} = 2\pi C r e^{-\frac{k}{2}(-\alpha r^2 + \frac{Q}{\alpha} + \frac{2m}{\alpha r})}.
\end{equation}
In Figure \ref{fig:ceq}, we plot the equatorial circumference for specific values of $\alpha$ and $m$.
{ Setting $k=k_0$, the derivative is given by
\begin{equation}
    \frac{d\mathcal{C}_{eq}}{dr} = 2\pi C e^{-\frac{k_0}{2}(-\alpha r^2 +\frac{Q}{\alpha} + \frac{2m}{\alpha r})} \left[ 1-\frac{k_0 r}{2} \left(-2\alpha r + \frac{Q'}{\alpha}-\frac{2m}{\alpha r^2}\right)\right].
\end{equation}
Setting this equal to zero gives
\begin{equation}
    2\alpha (1+2m) r^2 + 2\alpha m (1-2m)r +\frac{k_0}{2} = 0.
\end{equation}
Since $k_0$ is negative, the quadratic formula shows that this will have only one solution for $r>0$. We see that $\mathcal{C}'_{eq}(r=0)>0$.} The derivative at $r=1/\alpha$ is given by
\begin{equation}
    \frac{d\mathcal{C}_{\mathrm{eq}}}{dr}\bigg\vert_{r=\frac{1}{\alpha}} = 2\pi C e^{-\frac{k_0}{2}(2m-\frac{1}{\alpha})} \left[ 1-\frac{1}{4\alpha m} \ln\frac{1-2\alpha m}{1+2\alpha m}(\alpha m - 2) \right].
\end{equation}
It is easy to numerically verify that for $0<2\alpha m < 1$, this is always negative. Hence, the maximum will always occur for some $r<1/\alpha$. For $r=2m$ 
\begin{equation}\label{eq:diff-Ceq}
    \frac{d\mathcal{C}_{\mathrm{eq}}}{dr}\bigg\vert_{r=2m} = 2\pi C e^{-\frac{k_0}{2}}\left[  1+\frac{3\alpha m}{2} \ln \frac{1-2\alpha m}{1+2\alpha m} \right].
\end{equation}
Since in the above expression the parameters $\alpha$ and $m$ always appear as a product $\alpha m$, we numerically evaluate \eqref{eq:diff-Ceq} for $\alpha m \in (0,0.5)$, instead of varying them separately. We somewhat unexpectedly find that \eqref{eq:diff-Ceq} is only positive for $\alpha m < 0.36$. This implies that for $\alpha m<0.36$, the maximum occurs between the black hole and acceleration horizons and for $ 0.36 < \alpha m <0.5 $, the maximum circumference occurs \emph{inside} the black hole horizon.

One can also analyze the ``circumference'' (keeping $k$ for generality)
\begin{equation}
    \mathcal{C}_\theta(r) = 2\pi C \frac{\sqrt{P} \sin\theta}{1-\alpha r \cos\theta}  e^{-\frac{k}{2}(F+L)}r
\end{equation}
for values of $\theta\neq \pi/2$. To find the local maximum or minimum, one would again take the derivative with respect to $r$. In this case, however, the equation is not easily analyzed for general values of $\alpha$, $m$, and $\theta$.

For the polar circumference, $\mathcal{C}_{\mathrm{pol}}$, the line element 
\begin{equation}
    ds_{\text{pol}}=\frac{r e^{\frac{k}{2}(F-L)-\frac{k^2}{2}W}}{(1-\alpha r \cos\theta)\sqrt{1-2\alpha m \cos\theta}}d\theta.
\end{equation}
can be integrated numerically (it cannot be done analytically). A plot is shown in Figure \ref{fig:cpol}.

\subsubsection*{Radial Distance}
\begin{figure}[htb]
\centering
\begin{subfigure}[t]{0.24\textwidth}
\includegraphics[width=\textwidth]{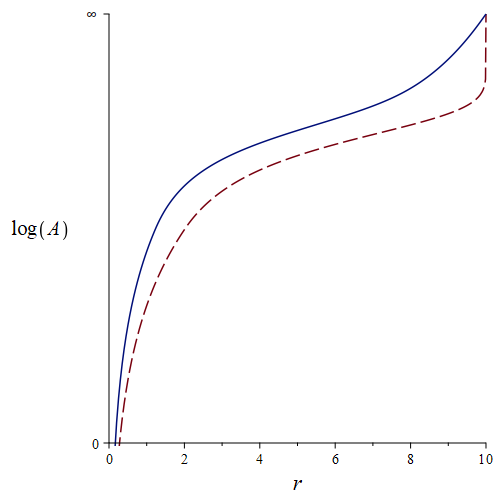}
\caption{Comparison of the surface areas of the C-metric and regularized C-metric.}
\label{fig:surface-area}
\end{subfigure}\hfill
\begin{subfigure}[t]{0.24\textwidth}
\includegraphics[width=\textwidth]{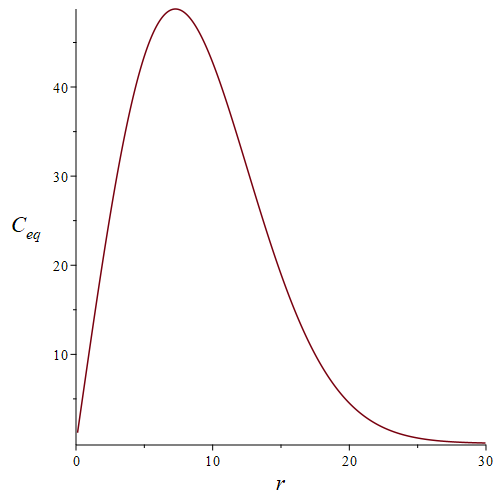}
\caption{$\mathcal{C}_{\mathrm{eq}}$ vs.~$r$.}
\label{fig:ceq}
\end{subfigure}\hfill
\begin{subfigure}[t]{0.24\textwidth}
\includegraphics[width=\textwidth]{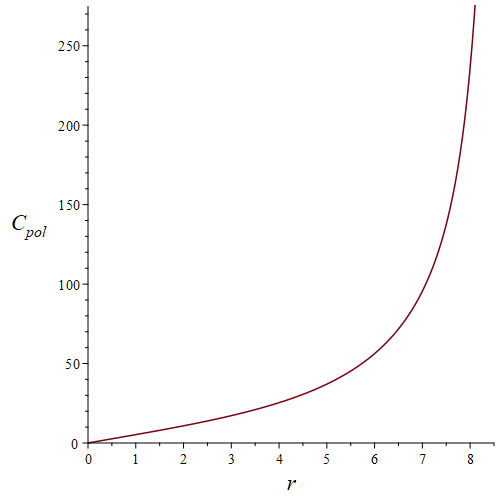}
\caption{$\mathcal{C}_{\mathrm{pol}}$ vs.~$r$.}
\label{fig:cpol}
\end{subfigure}\hfill
\begin{subfigure}[t]{0.24\textwidth}
\includegraphics[width=\textwidth]{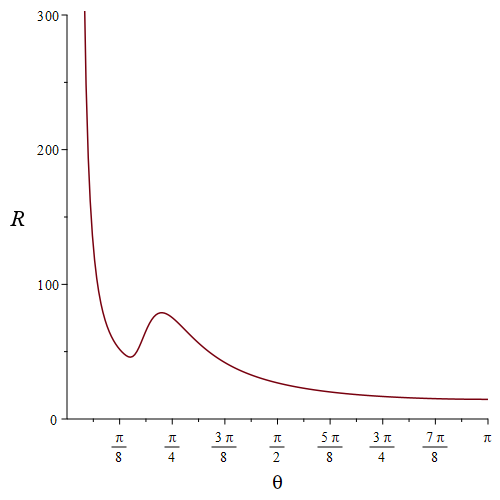}
\caption{Plot of radial distance $R$ vs.~$\theta$, from $r=2m$ to $1/\alpha$.}
\label{radial-distance}
\end{subfigure}
\caption{Figure \ref{fig:surface-area} shows the comparison of the surface areas of the regularized C-metric $(\mathrm{i.e.,~} k=k_0)$ vs.~the original C-metric $(\mathrm{i.e.,~} k=0)$ for $(m,\alpha) = (1,0.1)$, in logarithmic scale. Red (dash) represents the regularized C-metric and the blue (line) represents the original C-metric. Figures \ref{fig:ceq} and \ref{fig:cpol} plot the equatorial and polar circumferences, respectively, with $(m,\alpha) = (1, 0.1)$. Figure \ref{radial-distance} plots the radial distance $R$ from $r=2m$ to $1/\alpha$ vs.~$\theta$ for $(m,\alpha) = (1, 0.1)$.}
\label{circumference}
\end{figure}

The radial line element is given by
\begin{equation}
    ds = \frac{1}{1-\alpha r \cos\theta} \frac{e^{\frac{k}{2}(F-L)-\frac{k^2}{2}W}}{\sqrt{Q}} dr.
\end{equation}
Notice that for $\theta=0$, the distance from $r=2m$ to $1/\alpha$ is infinite. Figure \ref{radial-distance} shows the radial distance as a function of $\theta$.

\subsection{Embedding in \texorpdfstring{$\mathrm{E}^3$}{E3}} \label{embedding_section}
In their original work, Kinnersly and Walker showed that the space-like $t$-constant $r$-constant two-surfaces of the C-metric are embeddable in $\mathrm{E}^3$, for a particular periodicity of $\varphi$, i.e., for a particular $C$ \cite{Kinnersley_Walker}. Horizon deformation caused by the acceleration was subsequently studied in \cite{Farhoosh}. Here we will consider embeddability as function of $C$ for any $t$-constant $r$-constant two-surface. We then consider the black hole and the acceleration horizons of the regularized C-metric and find both to be embeddable in $\mathrm{E}^3$. 

\begin{figure}[htb]
\begin{subfigure}{0.24\textwidth}
    \centering
    \includegraphics[width=\textwidth]{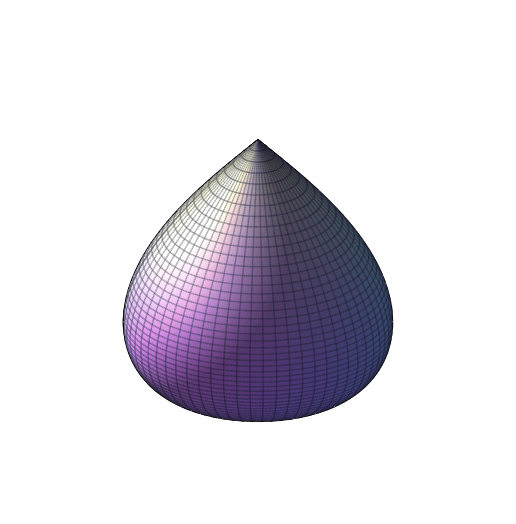}
    \caption{$C=1$, the sharp point on the top represents the conic singularity.}
    \label{fig:c=1}
\end{subfigure}\hfill
\begin{subfigure}{0.24\textwidth}
    \centering
    \includegraphics[width=\textwidth]{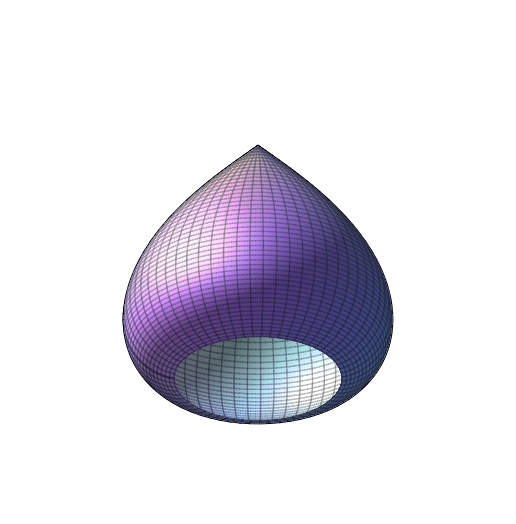}
    \caption{$C=1$, bottom hole shows the embeddability issue.}
    \label{fig:c=1-bottom}
\end{subfigure}\hfill
\begin{subfigure}{0.24\textwidth}
    \centering
    \includegraphics[width=\textwidth]{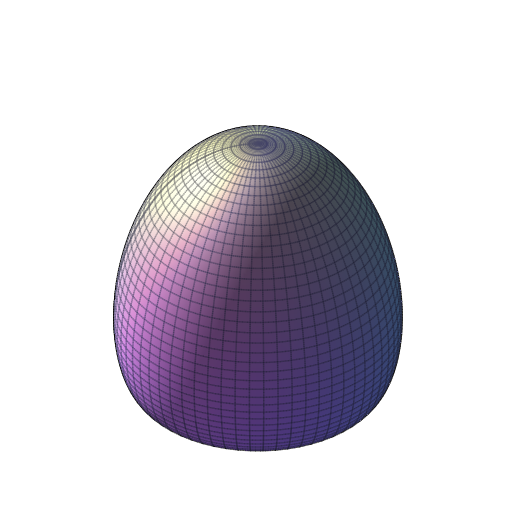}
    \caption{$C=\frac{1}{1-2\alpha m}$, with a regular north axis. It is \emph{not} completely embeddable.}
    \label{fig:c-top-reg}
\end{subfigure}\hfill
\begin{subfigure}{0.24\textwidth}
    \centering
    \includegraphics[width=\textwidth]{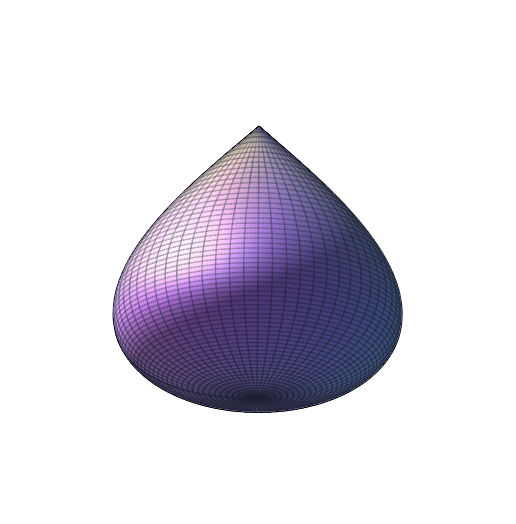}
    \caption{$C=\frac{1}{1+2\alpha m}$, with a regular south axis. The horizon is completely embeddable.}
    \label{fig:c-bottom-reg}
\end{subfigure}
\caption{C-metric black hole horizon showing conical singularities, embeddability issues and regularizing either side of the axis with appropriate choice of the range parameter $C$. Pictures are for $(m,\alpha) = (1,0.2)$. Horizons in Figures \ref{fig:c=1} (\ref{fig:c=1-bottom}) and \ref{fig:c-top-reg} are not entirely embeddable and will have a hole in the bottom. The horizon in Figure \ref{fig:c-bottom-reg} is completely embeddable, and is the diagram that is usually presented in the literature (for example, Figure 3 in \cite{Kinnersley_Walker} and Figure 3 in \cite{griffiths_krtous_podolsky}). These figures can also be compared with the line source diagrams in Figure \ref{conic_figure}.}
\label{embedding}
\end{figure}

\begin{figure}[htb]
\begin{subfigure}{0.24\textwidth}
    \centering
    \includegraphics[width=\textwidth]{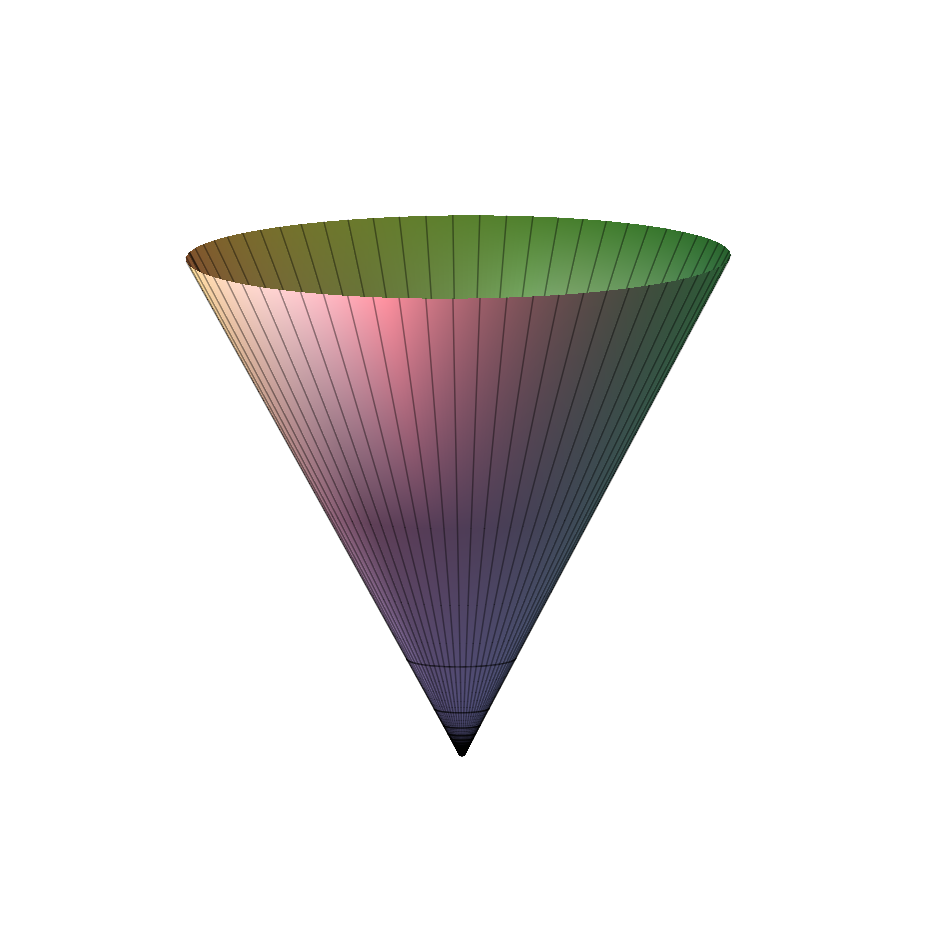}
    \caption{C-metric, $\alpha=0.2$}
    \label{fig:AH-C-alpha=0.2}
\end{subfigure}\hfill
\begin{subfigure}{0.24\textwidth}
    \centering
    \includegraphics[width=\textwidth]{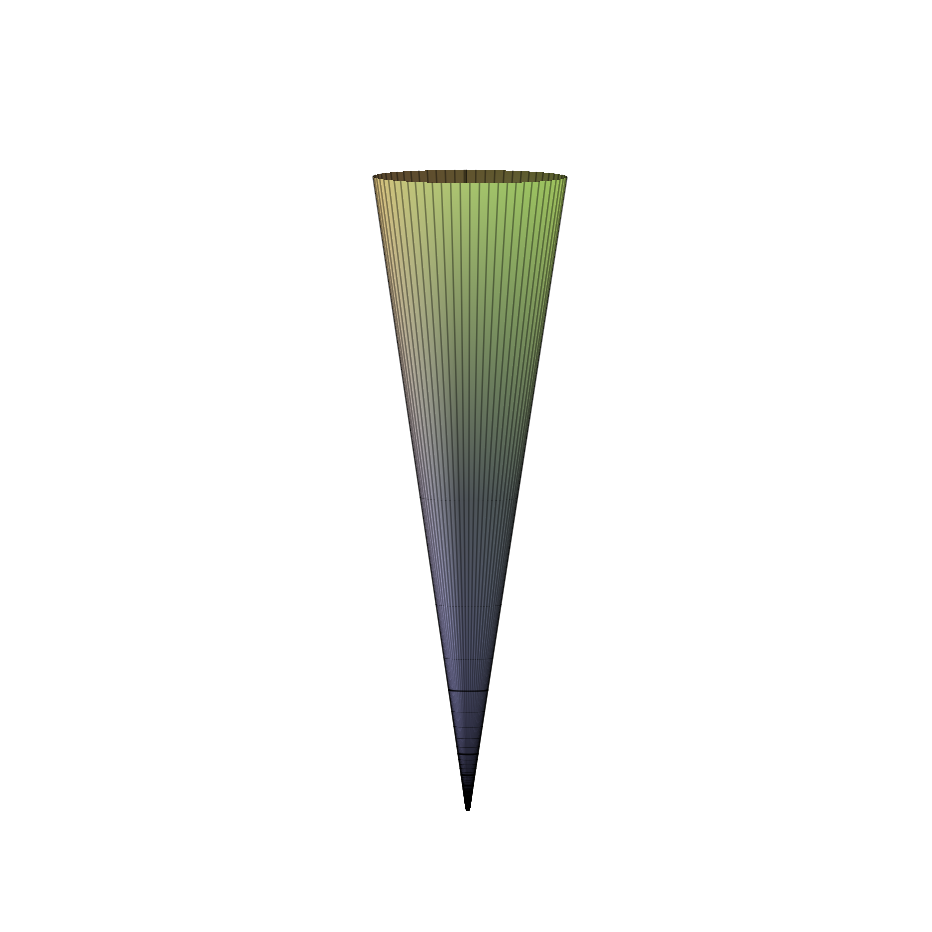}
    \caption{C-metric, $\alpha=0.4$}
    \label{fig:AH-C-alpha=0.4}
\end{subfigure}
\begin{subfigure}{0.24\textwidth}
    \centering
    \includegraphics[width=\textwidth]{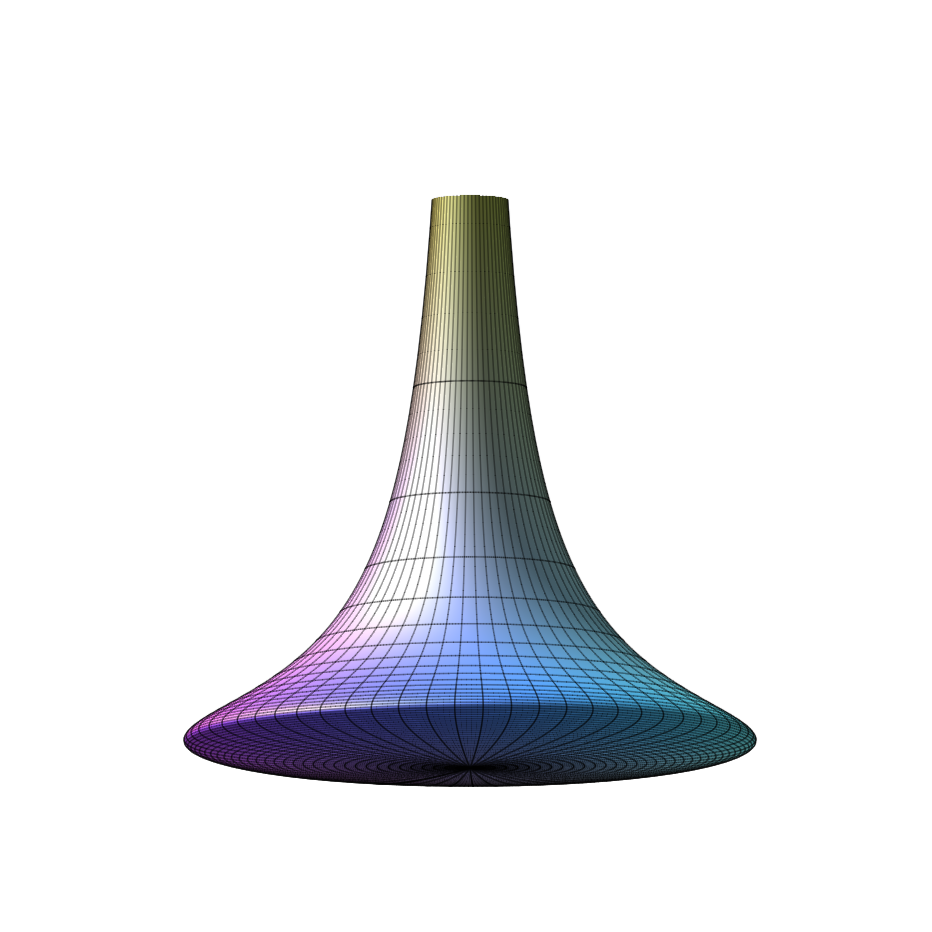}
    \caption{Regularized, $\alpha=0.2$}
    \label{fig:AH-GC-alpha=0.2}
\end{subfigure}\hfill
\begin{subfigure}{0.24\textwidth}
    \centering
    \includegraphics[width=\textwidth]{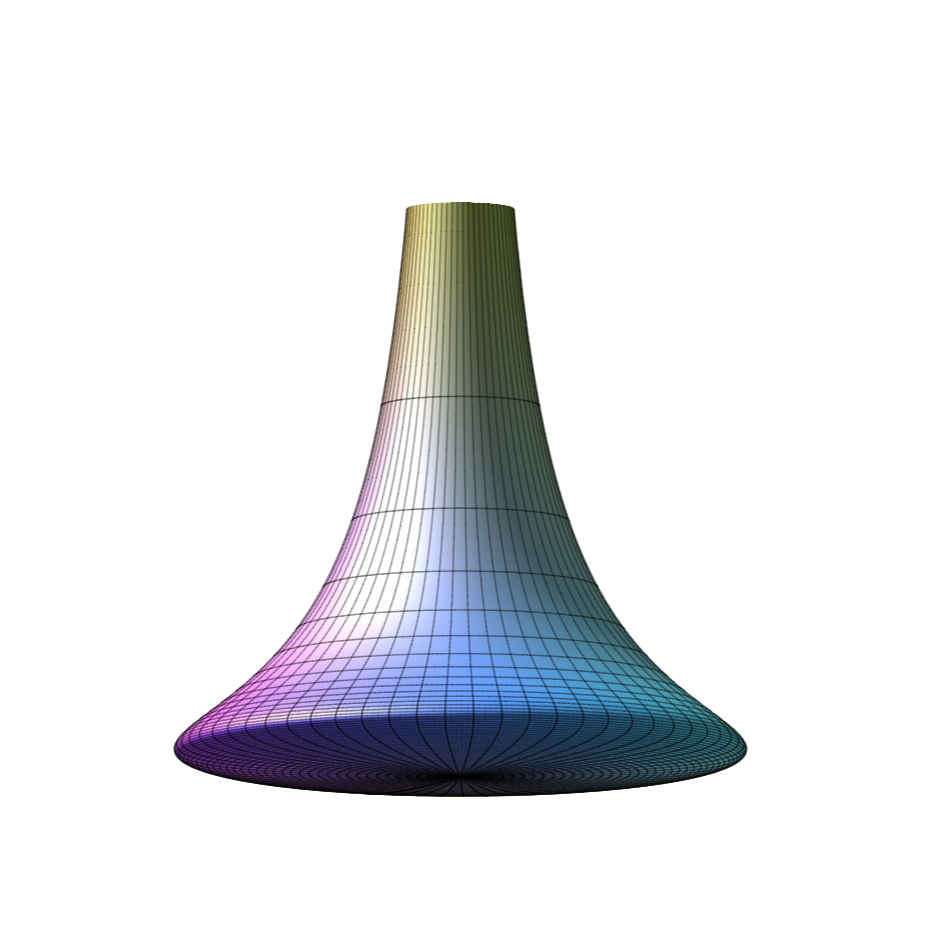}
    \caption{Regularized, $\alpha=0.4$}
    \label{fig:AH-GC-alpha=0.4}
\end{subfigure}
\caption{The acceleration horizons of the C-metric and the regularized C-metric for two values of $\alpha$, $m=1.0$. These surfaces are topologically $\mathbb{R}^2$.}
\label{fig:acceleration_horizons}
\end{figure}

For a two-dimensional metric (see, for example, \cite{Smarr, frolov2011})
\begin{equation}
ds^2 = f(\theta) d\theta^2 + C^2 g(\theta) d\phi^2~,
\end{equation}
where $\theta\in[0,\pi]$, $\phi\in[0,2\pi)$\footnote{$C$ has to be written explicitly in the metric so that the range of $\phi$ is $\phi\in[0,2\pi)$. This is done through $\varphi = C \phi$.} and $C$ is a positive constant, one can define the embedding functions as
\begin{equation}
x = F(\theta) \cos\phi~, \qquad y = F(\theta)\sin\phi~, \qquad z = G(\theta)~.
\end{equation}
Matching the metric coefficients gives
\begin{equation}
F^2 = C^2 g~, \qquad F'{^2} + G'{^2} = f~,
\end{equation}
which gives
\begin{equation}\label{g-solution}
G = \int \sqrt{f - F'{^2}} ~d\theta = \int \sqrt{f-C^2 \frac{g'{^2}}{4 g}} ~d\theta~.
\end{equation}
Thus, the condition for embeddability is 
\begin{equation}\label{embed-cond}
f - C^2 \frac{g'{^2}}{4 g} \ge 0~.
\end{equation}
For the black hole horizon of the (original) C-metric, the condition \eqref{embed-cond} becomes 
\begin{equation}\label{embed-cond1}
C^2\left[\cos\theta - \alpha m (1+\cos^2\theta)\right]^2 \leq 1~.
\end{equation}
For $C=1$, since the value of the left-hand side at $\theta=\pi$ is $(1+2\alpha m)^2$ (which happens to be the maximum of the expression), there would always exist a $\theta\in [0,\pi]$, above which the horizon is not embeddable\footnote{We consider partial embeddability in the sense that \eqref{embed-cond} can be satisfied for a sub-range of $\theta$, not necessarily the entire range.}. One can choose a sufficiently small value of $C <1$ so that the maximum value of the expression is less than one (for a particular $\alpha m$). For $C=1/(1+2\alpha m)$, the embeddability condition is satisfied \footnote{With  $C<1/(1+2\alpha m)$ one can also satisfy \eqref{embed-cond1}, but this would result in neither pole being regular.} (this is equivalent to the value that Kinnersley and Walker considered \cite{Kinnersley_Walker}). The condition for the acceleration horizon is given by
\begin{equation}
\frac{C^{2}}{\left(1-\cos\theta\right)^{2}} \left[ -1 + \cos\theta + \alpha m (1 + \cos\theta) - \alpha m \cos^2\theta (3-\cos\theta) \right]^{2} \leq 1~.
\end{equation}
One can easily check that the embeddability condition is satisfied for the acceleration horizon, for the same value of $C$.
Figure \ref{embedding} shows the embedding diagrams for the horizon of the C-metric for different choices of $C$. Figures \ref{fig:AH-C-alpha=0.2} and \ref{fig:AH-C-alpha=0.4} show the embedding diagrams for the acceleration horizon of the C-metric.

\begin{figure}[htb]
\begin{subfigure}{0.24\textwidth}
    \centering
    \includegraphics[width=\textwidth]{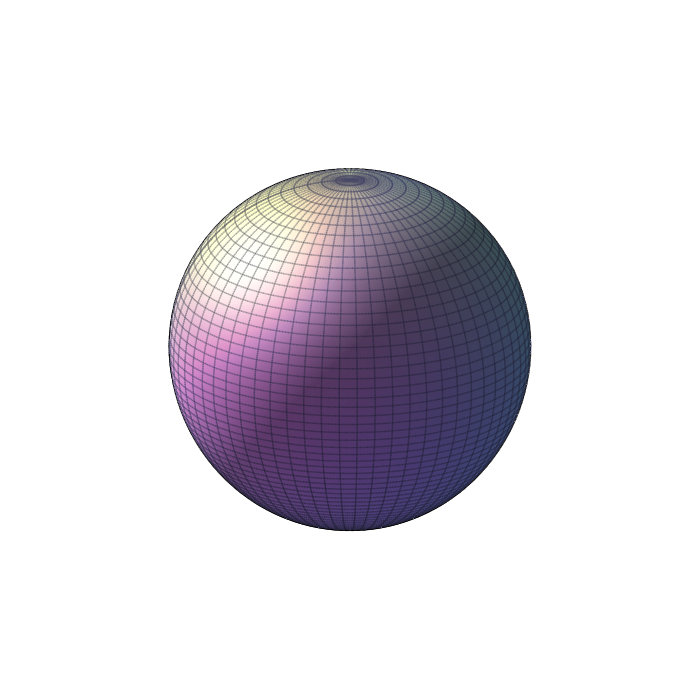}
    \caption{$\alpha=0.2$}
\end{subfigure}\hfill
\begin{subfigure}{0.24\textwidth}
    \centering
    \includegraphics[width=\textwidth]{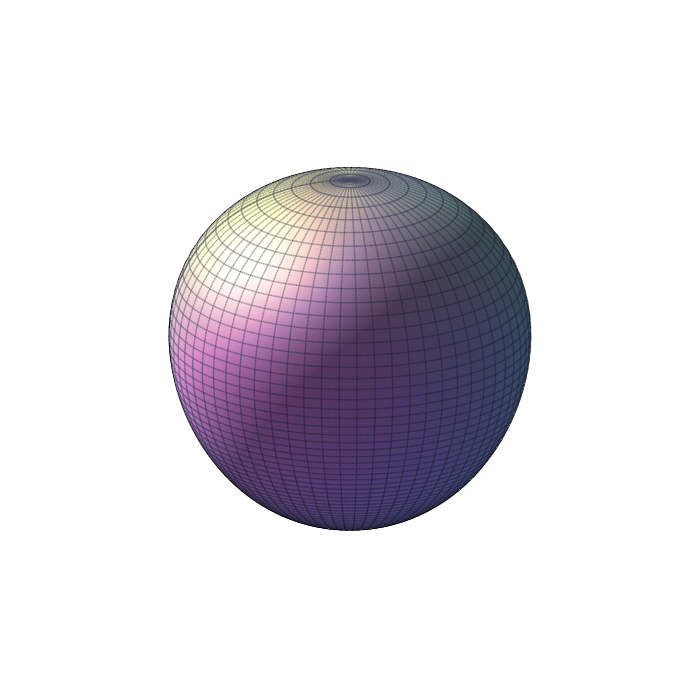}
    \caption{$\alpha=0.3$}
\end{subfigure}
\begin{subfigure}{0.24\textwidth}
    \centering
    \includegraphics[width=\textwidth]{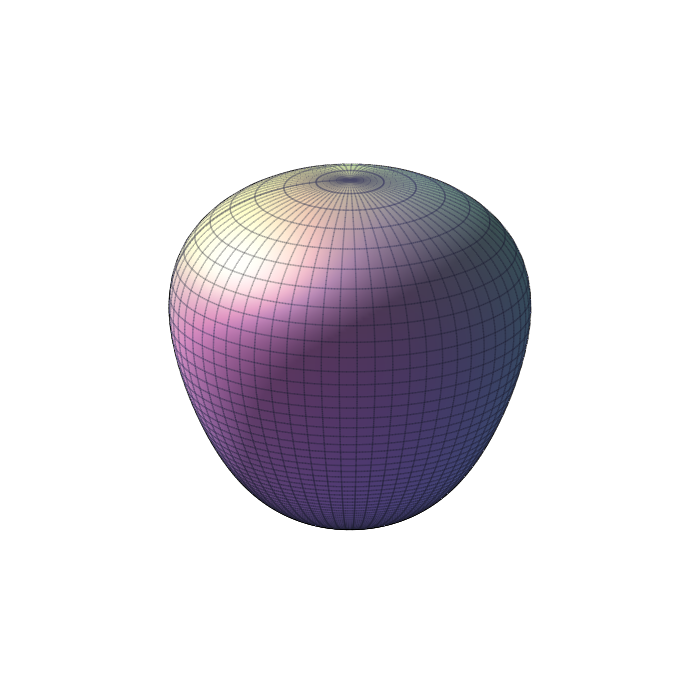}
    \caption{$\alpha=0.4$}
\end{subfigure}\hfill
\begin{subfigure}{0.24\textwidth}
    \centering
    \includegraphics[width=\textwidth]{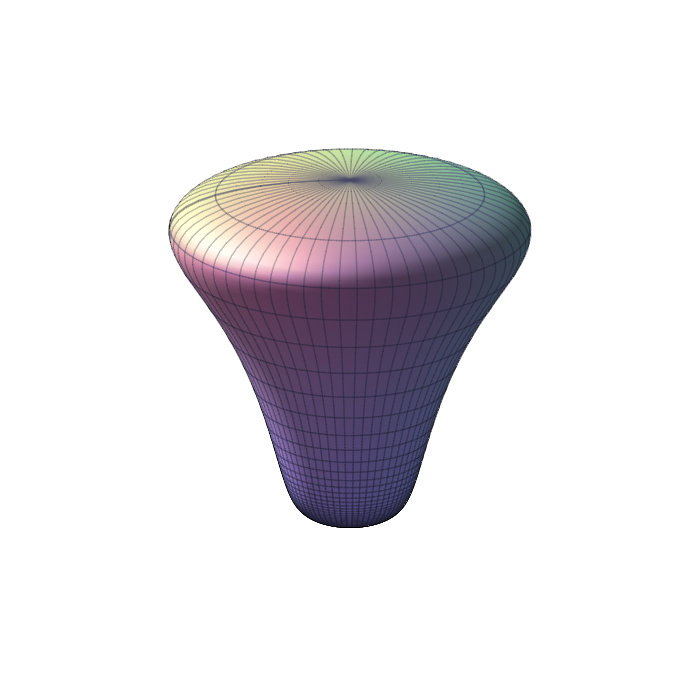}
    \caption{$\alpha=0.49$}
\end{subfigure}
\caption{The black hole horizon of the generalized C-metric for various values of $\alpha$, $m=1$. The value of $k$ is given by equation \eqref{eqn:k-value}. Both axes are regular.}
\label{ernst_embedding}
\end{figure}

In contrast, the black hole horizon of the regularized C-metric (for the values of $k$ and $C$ given by \eqref{eqn:k-value}) is always embeddable. The embeddability condition for a $r$-constant and $t$-constant surface of the generalized C-metric is given by
\begin{multline}\label{embed-condition}
(1-\alpha r \cos\theta)^2 ~-~ C^2e^{k^2 W-2kF} \Biggl[ m \alpha \cos^2\theta \left(\alpha r \cos\theta - 3\right)  + \cos\theta \left(r m \alpha^{2} + 1\right) -\alpha \left(r-m\right) \\ - \frac{1}{2} k\sin\theta P (1-\alpha r \cos\theta) \left( F' + L' \right) \Biggr]^2 \geq 0,
\end{multline}
where $P$ and $C$ are as in \eqref{c-metric-func} and \eqref{eqn:k-value}, respectively, and prime denotes the derivative with respect to $\theta$. Both quantities on the left-hand side in equation \eqref{embed-condition} are positive; their difference does not need to be positive for all $\theta$. In particular, for $r=2m$ the condition becomes
\begin{equation}
C^2 e^{-2 k F_{(r=2m)}} \left[ \frac{km\sin^2\theta}{PC^2} + \cos\theta -\alpha m \left(1+\cos^2\theta \right)  \right]^2 \leq 1~.
\end{equation}
The maximum value of the expression on the left-hand side is $1$, and occurs at $\theta = 0$. One can check numerically that the inequality is satisfied for all other values of $\theta\in [0,\pi]$ and $0<2\alpha m< 1$. Thus, the horizons of the regularized C-metric are always embeddable. Similarly, one can check that the acceleration horizon is embeddable for $\theta\in(0,\pi]$. Figure \ref{ernst_embedding} show embedding figures for some values of the parameters $\alpha$ and $m$. They also show regularity of the axis. Figures \ref{fig:AH-GC-alpha=0.2} and \ref{fig:AH-GC-alpha=0.4} show the embedding diagrams for the acceleration horizon of the regularized C-metric.

\section{Weyl coordinates and Line sources}\label{linesource}
In this section, we investigate the effect of regularization on the line sources of the C-metric. The best-suited coordinates for this purpose are the Weyl coordinates.

\subsection{Generalized C-metric in Weyl coordinates}
The effect of Ernst's prescription on solutions written in Weyl coordinates takes a simpler form \cite{stephani_exact_sol,Akbar_MacCallum}. The functions $u,\nu$ change as follows
\begin{equation}
u \to u + kz~, \qquad  \nu \to \nu + kF - \frac{k^2}{2}\rho^2~,
\end{equation}
where $k$ is a new parameter and $F$ is a new auxiliary function obtained by solving the differential equation
\begin{equation}
\nabla F = 2\mathrm{i} \rho \nabla u~,\quad \text{where} \quad \nabla = \partial_\rho + \mathrm{i}\partial_z~.
\end{equation}
Therefore, the generalized C-metric in Weyl coordinates is given by
\begin{equation}
ds^2 = -e^{2u}dt^2 + e^{2\nu-2u}(d\rho^2 + dz^2) + \rho^2 e^{-2u} d\phi^2~,
\end{equation}
where the functions $u,\nu$ are given by
\begin{align}
e^{2u} &= \alpha \frac{R_1+R_2-2m}{R_1+R_2+2m} \left( R_3 + z + \frac{1}{2\alpha} \right) e^{2kz}~,\\
e^{-2u+2\nu} &= \frac{\left[ (1-2\alpha m)R_1 + (1+2\alpha m)R_2 + 4\alpha m R_3 \right]^2}{8\alpha \left(1-4\alpha^2 m^2\right)^2 R_1 R_2 R_3} e^{2kF - k^2 \rho^2 - 2kz}~,
\end{align}
with
\begin{gather}\label{r-def}
R_i \equiv \sqrt{\rho^2 + (z-z_i)^2}~, \qquad \qquad z_1 = m~,~~ z_2=-m~,~~ z_3 = -\frac{1}{2\alpha}~,\\ 
F = z - R_1 + R_2 - R_3~.
\end{gather}

\subsubsection*{Line sources of the generalized C-metric}
The modified functions $u$ and $\nu$ for the generalized C-metric are given by (see equations \eqref{eq:modified-u} - \eqref{eq:modified-nu})
\begin{equation}
e^{2u} = \frac{1}{C^2} \alpha e^{kz} \frac{R_1+R_2-2m}{R_1+R_2+2m}\left( R_3 + z + \frac{1}{2\alpha} \right)~,
\end{equation}
\begin{equation}
e^{2\nu} = \frac{\alpha}{C^2} \frac{R_1+R_2-2m}{R_1+R_2+2m}\left( R_3 + z + \frac{1}{2\alpha} \right) \frac{\bigl[ (1-2\alpha m)R_1 + (1+2\alpha m)R_2 + 4\alpha m R_3 \bigr]^2}{8\alpha (1-4\alpha^2m^2)^2R_1R_2R_3}e^{2kF - k^2\rho^2}.
\end{equation}
{ From equation \eqref{r-def},} $F(\rho =0, z)$ for either $z>m$ or $-\frac{1}{2\alpha}<z<-m$ is given by 
\begin{equation}
F(\rho = 0, z) = z + \sqrt{(z+m)^2} - \sqrt{(z-m)^2} - (z+\frac{1}{2\alpha}){ .}
\end{equation}
{ The function $e^{\nu_0}$} for $z>m$ takes the form
\begin{equation}\label{eq:nu-0-1}
e^{\nu_0} = \sqrt{\frac{1+2\alpha m}{1-2\alpha m}} e^{k(2m-\frac{1}{2\alpha})}~,
\end{equation}
{  and} for $-\frac{1}{2\alpha}<z<-m$, takes the form
\begin{equation}\label{eq:nu-0-2}
e^{\nu_0} =\sqrt{ \frac{1-2\alpha m}{1+2\alpha m} } e^{k(-2m-\frac{1}{2\alpha})}~.
\end{equation}
{ The value of $k$ which regularizes the axis can be found by equating the expressions for $e^{\nu_0}$ in equations \eqref{eq:nu-0-1} and \eqref{eq:nu-0-2}:}
\begin{equation}
k = \frac{1}{4m}\ln\frac{1-2\alpha m}{1+2\alpha m}~.
\end{equation}
Using equation \eqref{eq:linesource}, the energy momentum density on the axis is computed to be
\begin{equation}
    \mathcal{L}^2_2 = \mathcal{L}^4_4 = 0~,
\end{equation}
which essentially shows that there is no line source in the regularized C-metric. This gives more weight to Ernst's hypothesis that conical singularities, at least for spacetimes describing accelerating sources, appear because of the lack of information on the source causing the acceleration. In the case of the generalized C-metric the external field parameter ``encodes'' the source of acceleration, and the line source vanishes. Note that we excluded the $-m<z<m$ range of the $z$-coordinate above as this region corresponds to the black hole horizon. Line sources are considered on the actual symmetry axis of the spacetime.

\subsubsection*{Newtonian Interpretation}\label{new.inter.}
\begin{figure}[htb]
\begin{subfigure}[t]{0.3\textwidth}
    \centering
    \includegraphics[width=\textwidth]{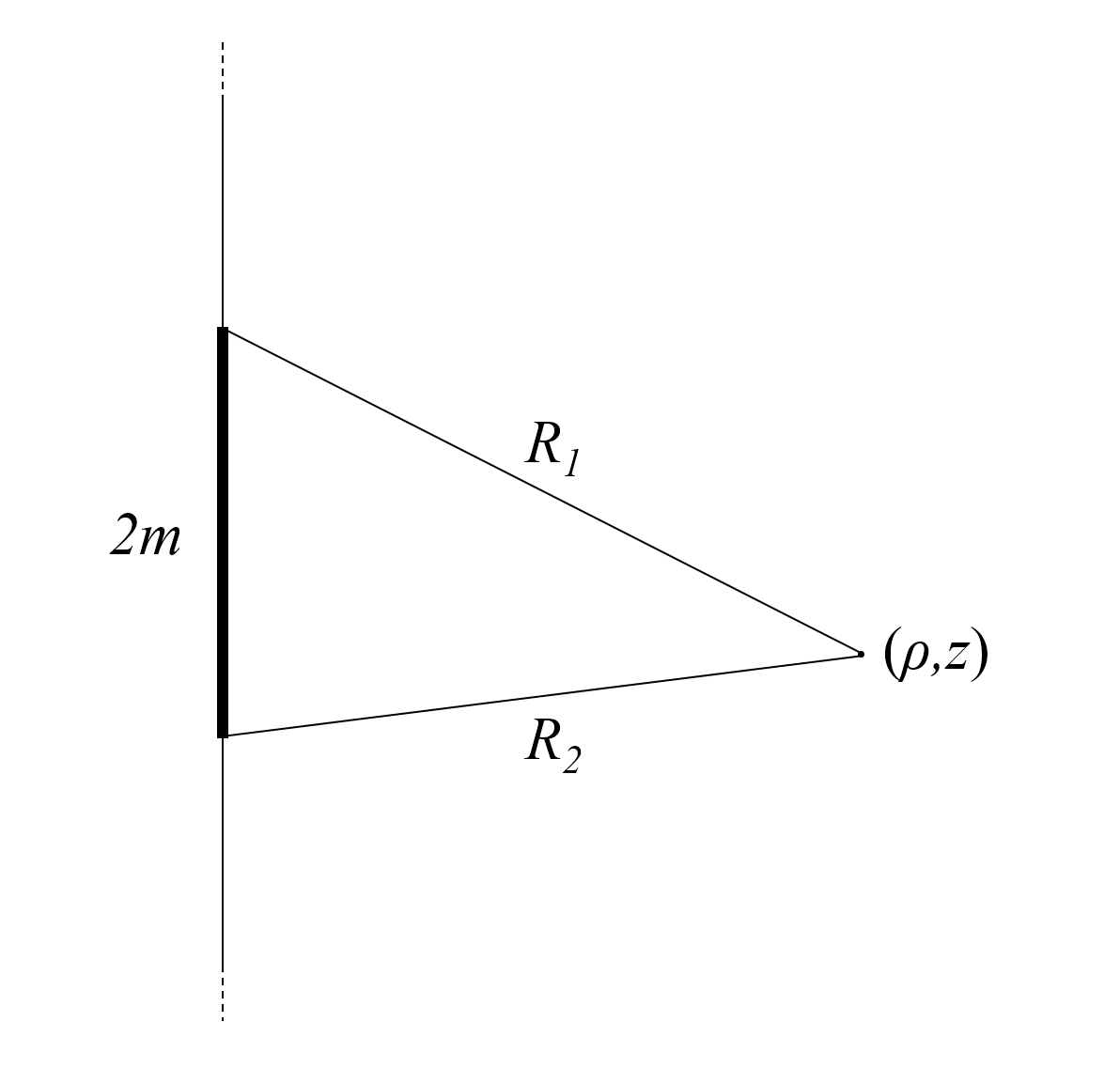}
    \caption{Schwarzschild in Weyl coordinates (finite line mass)}
\end{subfigure}\hfill
\begin{subfigure}[t]{0.3\textwidth}
    \centering
    \includegraphics[width=\textwidth]{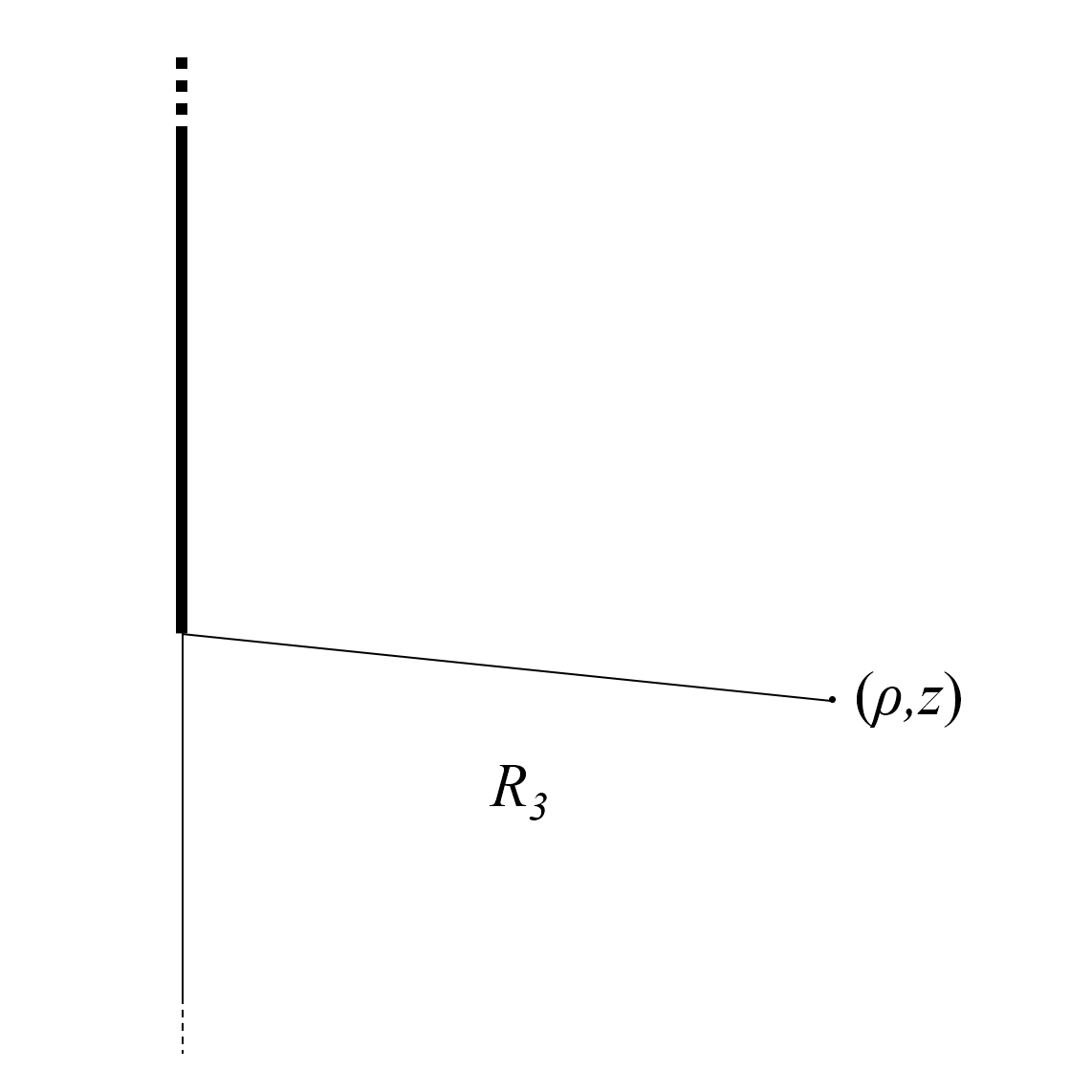}
    \caption{Accelerating particle (semi-infinite line mass)}
\end{subfigure}\hfill
\begin{subfigure}[t]{0.3\textwidth}
    \centering
    \includegraphics[width=\textwidth]{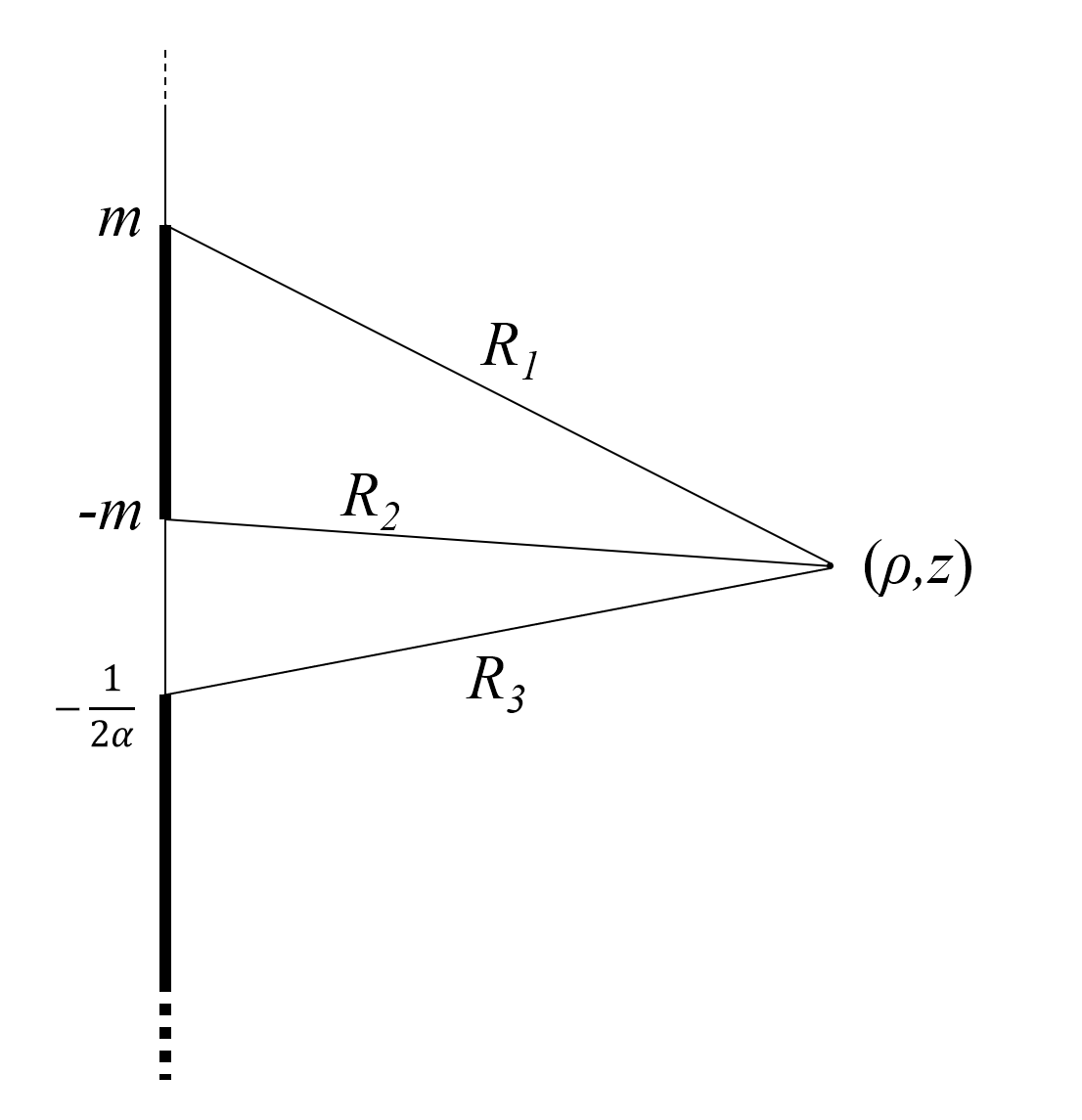}
    \caption{C-metric (finite line mass and a semi-infinite line mass)}
\end{subfigure}
\caption{Newtonian representation of the C-metric is a superposition of a finite line mass and a semi-infinite line mass.}
\label{Newtonian}
\end{figure}
The function $u$ in the Weyl metrics satisfies Laplace's equation in three dimensions and is therefore interpreted as a natural analogue of the Newtonian potential \cite{stephani_exact_sol}. The potential describing the C-metric is thus given the interpretation as a superposition of a finite line mass (describing the black hole) and a semi-infinite line mass (SILM, describing the acceleration horizon), as shown in Figure \ref{Newtonian} \cite{Bonnor}. For the generalized C-metric, the new potential is given by $u = u_0 + kz$. Thus, the generalization can be thought of as the above system being placed in a ``pseudo-uniform gravitational field" whose strength is $kz$ \cite{Bonnor88}. This is a purely relativistic effect since no such potential can originate from Newtonian sources\footnote{ The closest Newtonian source is that of an infinite sheet of uniform mass density $\sigma$, whose potential is given by $2\pi\sigma |z|$}.

\subsection{Two black hole interpretation}\label{bs-section}
\begin{figure}[htb]
    \centering
    \includegraphics[width=0.4\textwidth]{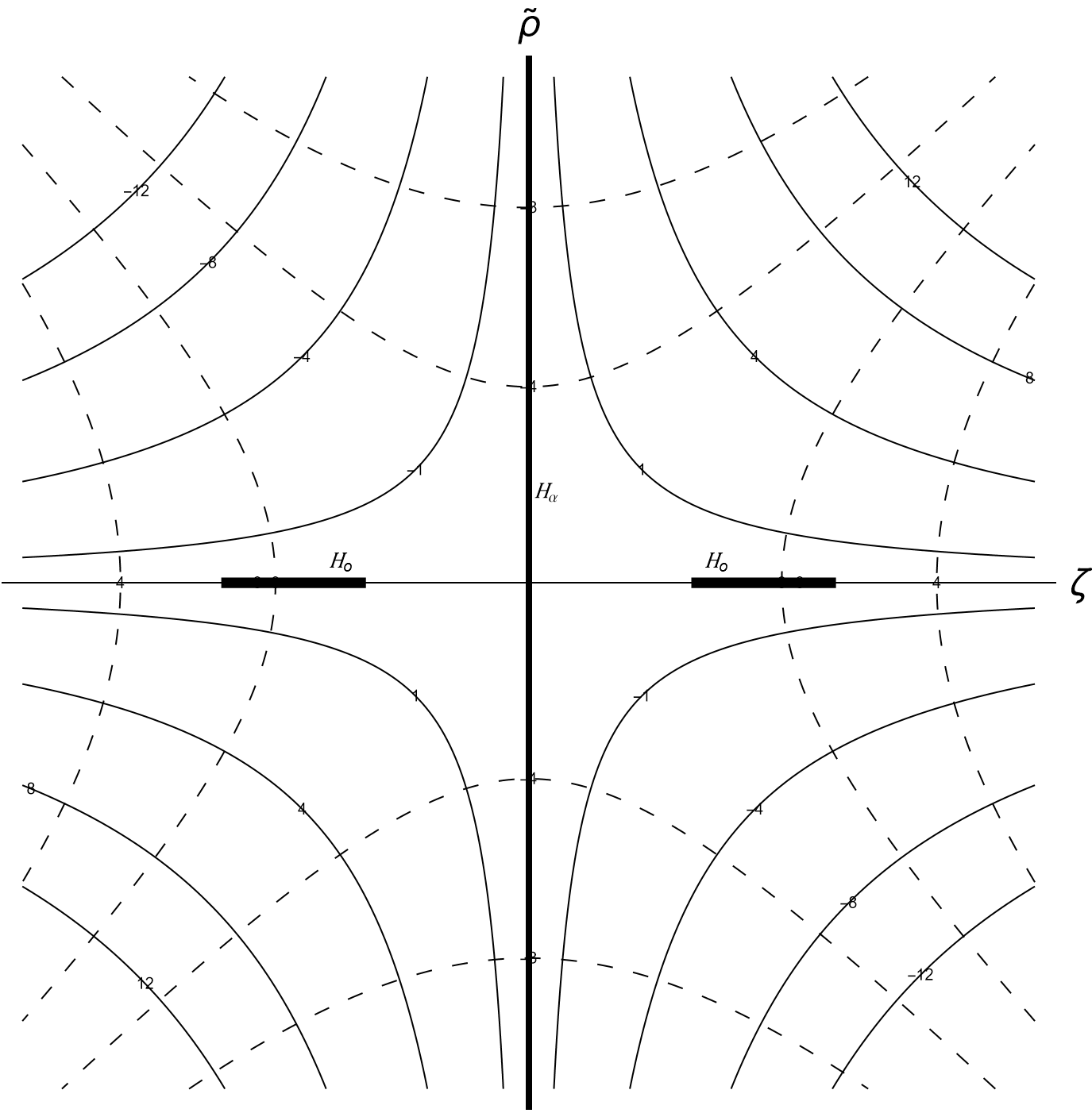}
    \caption{Weyl coordinate curves $\rho=constant$ (lines) and $z=constant$ (dashes), on top of boost-rotation symmetric coordinates. The two thick lines represent the horizons of the two black holes. The line $\zeta=0$ is the acceleration horizon. Weyl coordinates only cover the right half of the figure.}
    \label{boost-rotation-cmetric}
\end{figure}

\begin{figure}[htb]
\begin{subfigure}{0.24\textwidth}
    \centering
    \includegraphics[width=\textwidth]{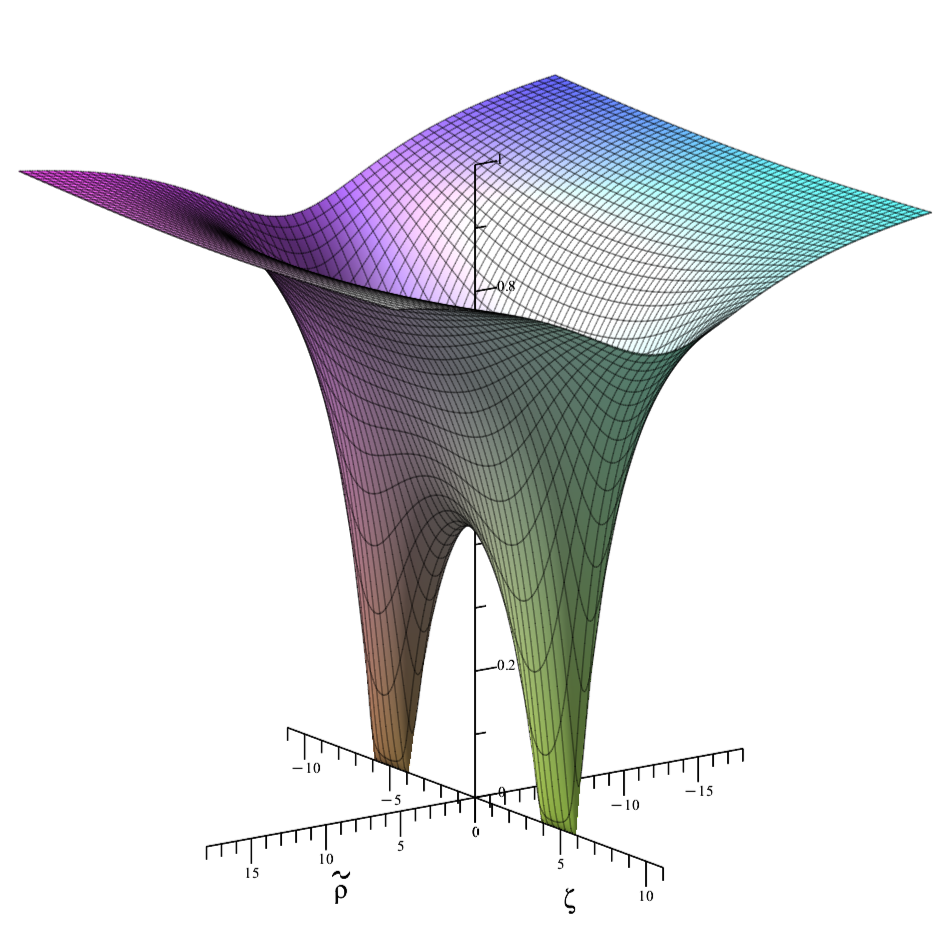}
    \caption{C-metric, $k=0$}
    \label{c-metric-bscoordinates}
\end{subfigure}\hfill
\begin{subfigure}{0.24\textwidth}
    \centering
    \includegraphics[width=\textwidth]{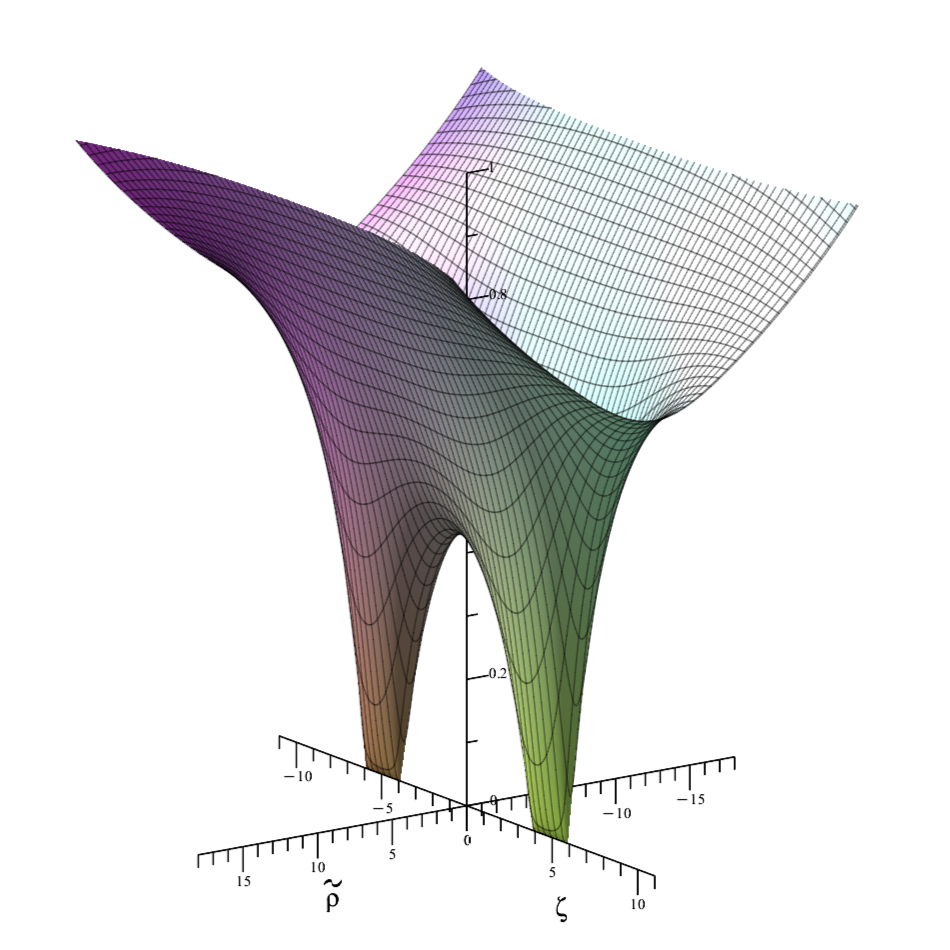}
    \caption{$k = -0.005$}
\end{subfigure}
\begin{subfigure}{0.24\textwidth}
    \centering
    \includegraphics[width=\textwidth]{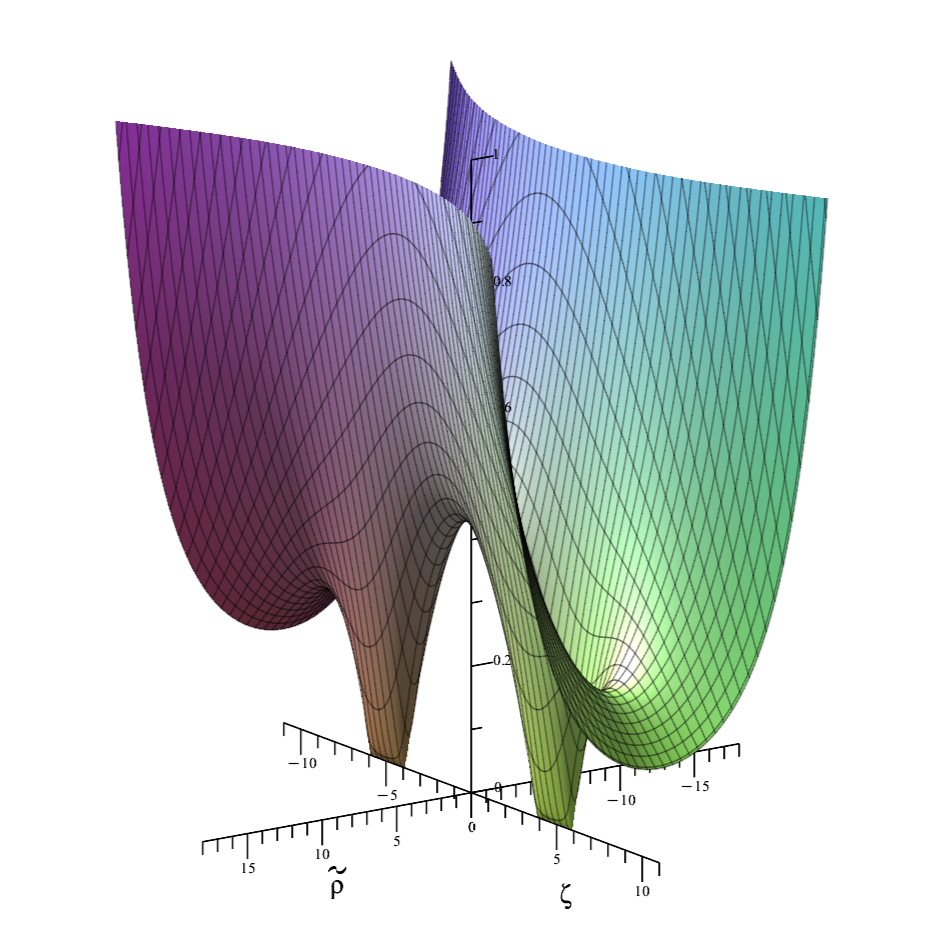}
    \caption{$k=-0.07$}
\end{subfigure}\hfill
\begin{subfigure}{0.24\textwidth}
    \centering
    \includegraphics[width=\textwidth]{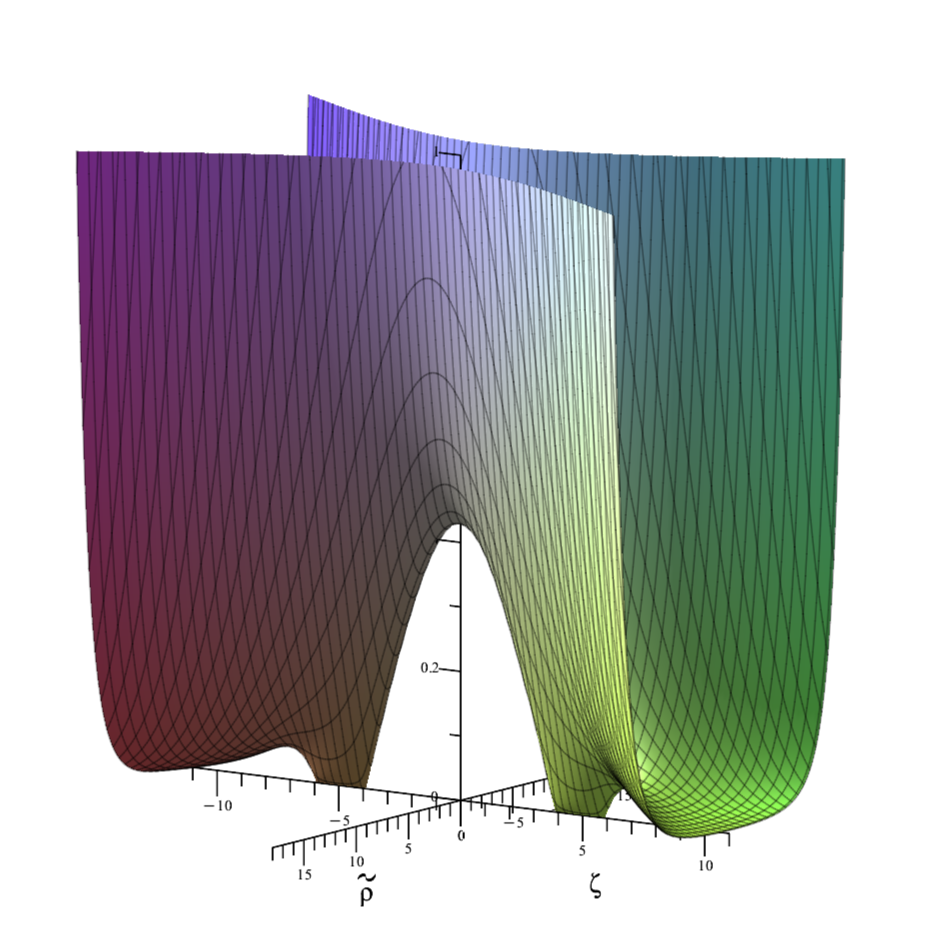}
    \caption{$k=-0.211$}
\end{subfigure}
\caption{The metric function $e^\mu$ is plotted to show the location of the two black holes for various values of $k$. $(m,\alpha) = (1,0.2)$. Figure $\mathrm{(d)}$ represents the regularized C-metric.}
\label{c-boostrotation}
\end{figure}

As mentioned earlier, the C-metric in boost-rotation symmetric form is one way of seeing that the spacetime represents two causally separated black holes accelerating away from each other. The same can be shown for the generalized C-metric by applying the hyperbolic parabolic orthogonal transformations one applies to the C-metric \cite{BicakSchmidt}:
\begin{equation}
    z = \frac{\alpha}{2}\left(\zeta^2 - \Tilde{\rho}^2\right) - \frac{1}{2\alpha}~, \qquad \rho = \alpha \zeta \tilde{\rho} \qquad \text{and} \qquad t = \frac{\tau}{\alpha}~,
\end{equation}
which gives
\begin{equation}\label{bs-form}
    ds^2 = -e^\mu \zeta^2 d\tau^2 + e^\lambda\left( d\zeta^2 + d\Tilde{\rho}^2 \right) + e^{-\mu}\Tilde{\rho}^2 d\varphi^2~,
\end{equation}
where
\begin{gather}
    e^\mu = \frac{R_1+R_2-2m}{R_1+R_2+2m} e^{k\left( \alpha(\zeta^2-\Tilde{\rho}^2) - \frac{1}{\alpha} \right)}~, \\
    e^\lambda = \frac{\left[ (1-2\alpha m)R_1 + (1+2\alpha m)R_2 + 4\alpha m R_3 \right]^2}{4 \left(1-4\alpha^2 m^2\right)^2 R_1 R_2} e^{k(-2R_1+2R_2-\alpha(\tilde{\rho}^2+\zeta^2)) - k^2\alpha^2\zeta^2\Tilde{\rho}^2 }~,
\end{gather}
and the $R_i$'s are the same functions as defined in equation \eqref{r-def}, now as functions of $\tilde{\rho}$ and $\zeta$.  Note that while the Weyl coordinates only cover the positive-$\zeta$ side, the boost-rotation symmetric coordinates show both sides, see Figure \ref{boost-rotation-cmetric}.
Figure \ref{c-boostrotation} shows how the metric function $e^\mu$ changes for different values of $k$. As can be seen in the figures, the zeros of the metric coefficient do not change (which represent the black hole horizons). One can also infer how the strength of the external field causes acceleration. 

\subsubsection*{Extending into region I}
Although the above form of the metric shows two black holes, it only covers the two static regions (region II) around these black holes, see Figure \ref{CoordPlot}. With a second copy of the coordinates, and with the transformations $T=\pm\zeta\sinh\tau,~ Z=\pm\zeta\cosh\tau$ in the two static regions (region II) and $T=\pm\zeta\cosh\tau,~ Z=\pm\zeta\sinh\tau$ in the two regions outside the acceleration horizons (region I), the metric \eqref{bs-form} can be explicitly written in the boost-rotation symmetric form (in coordinates that cover all regions I and II)
\begin{equation}
ds^2 = \frac{1}{Z^2-T^2}\left[-e^\mu\left(ZdT-TdZ\right)^2 + e^\lambda\left( ZdZ - TdT \right)^2 \right] + e^\lambda d\tilde{\rho}^2 + e^{-\mu}\tilde{\rho}^2 d\varphi^2~,
\end{equation}
in which  $T=\pm Z$ are the two acceleration horizons\footnote{As in the case of the original C-metric, the black hole horizons are not represented in this form.}.

\subsubsection{Ernst's prescription as a symmetry of the wave equation}
Here we note how Ernst's prescription translates to boost-rotation symmetric coordinates. In regions $Z^2>T^2$, the function $\mu$ in boost-rotation symmetric spacetimes satisfies the two-dimensional wave equation \cite{BicakSchmidt}
\begin{equation}\label{wave-equation}
    \mu_{,\Tilde{\rho}\Tilde{\rho}} + \mu_{,\zeta\zeta} + \frac{1}{\tilde{\rho}}~\mu_{,\tilde{\rho}} + \frac{1}{\zeta}~\mu_{,\zeta} = 0~
\end{equation}
which clearly has the symmetry
\begin{equation}
\mu \to \mu_0+\kappa(\zeta^2-\Tilde{\rho}^2)~, \qquad \kappa\in \mathbb{R}~.
\end{equation}
The other metric function $\lambda$ satisfies the two differential equations
\begin{align}
\left( \tilde{\rho}^2 + \zeta^2 \right)\lambda_{,\tilde{\rho}} &= \frac{1}{2}\tilde{\rho}\zeta^2 \left(\mu_{,\tilde{\rho}}^2 - \mu_{,\zeta}^2 \right)  + \tilde{\rho}^2\zeta \mu_{,\tilde{\rho}}\mu_{,\zeta} + \left( \tilde{\rho}^2 - \zeta^2 \right)\mu_{,\tilde{\rho}} - 2\tilde{\rho}\zeta \mu_{,\zeta}~,\\
\left( \tilde{\rho}^2 + \zeta^2 \right) \lambda_{,\zeta} &= -\frac{1}{2}\tilde{\rho}^2 \zeta \left(\mu_{,\tilde{\rho}}^2 - \mu_{,\zeta}^2 \right)  + \tilde{\rho} \zeta^2 \mu_{,\tilde{\rho}}\mu_{,\zeta} + \left( \tilde{\rho}^2 - \zeta^2 \right)\mu_{,\zeta} + 2\tilde{\rho}\zeta \mu_{,\tilde{\rho}}~.
\end{align}
Hence, Ernst's solution-generating technique takes the following form
\begin{equation}
    \mu \to \mu_0+\kappa(\zeta^2-\Tilde{\rho}^2)~, \qquad \lambda \to \lambda_0 + 2\kappa F -\frac{\kappa}{2}\left(\zeta^2+\Tilde{\rho}^2\right) - \kappa^2 \zeta^2 \Tilde{\rho}^2~,
\end{equation}
where $F$ solves
\begin{equation}
    \nabla F = \mathrm{i}\tilde{\rho}\zeta\nabla \mu_0~, \qquad \text{where} \qquad \nabla = \frac{\partial}{\partial \Tilde{\rho}} + \mathrm{i}\frac{\partial}{\partial \zeta}~.
\end{equation}
In the regions $T^2>Z^2$, the wave equation is slightly modified, and the above results apply after a transformation $(\zeta,\tau) \to (\mathrm{i}\tau, \zeta)$.

\section{Asymptotic structure}\label{asym_str}

\subsection{Conformal Diagrams}\label{conformal_diagrams}
By introducing an external field, we cure the conical singularities; this also distorts the  axisymmetric $r=\text{const}$ surfaces while keeping their spherical topology. In a sense, the regularized C-metric is a distortion of the original C-metric. However, as we will see below, there will be critical differences not only in the form of new curvature singularities but also in the topology of $\mathscr{I}$ which will no longer be $\mathbb{R}\times \mathrm{S}^2$.

The conformal diagrams of the original C-metric were first found by \cite{Kinnersley_Walker}. Later, this was also done by \cite{griffiths_krtous_podolsky} using Hong-Teo coordinates. To find the conformal diagrams of the regularized C-metric, we follow the methods of the latter. Starting with the Hong-Teo coordinates, they first find a tortoise coordinate, then set $\theta, \phi$ constant, and then find Kruskal-type coordinates that allows the metric to cross the horizons $r=2m$ or $r=1/\alpha$. Our approach will be the same, except that we must first set $\theta, \phi$ as constant and then find a tortoise coordinate for the resulting $2$-dimensional metric. However, this method, as in \cite{griffiths_krtous_podolsky}, obtains the conformal diagram of $\theta, \phi$ constant surfaces which we will then combine to obtain the $3$-dimensional conformal diagram.

Setting $\theta, \phi$ constant, the resulting 2-metric will be
\begin{equation}
    ds_2^2 = \frac{1}{(1-\alpha r \cos\theta)^2} \left( -Q(r) e^{k_0 (F+L)}dt^2 + \frac{e^{k_0 (F-L)-k_0^2 W}}{Q(r)} dr^2 \right), \hspace{2em} r\leq \frac{1}{\alpha}.
\end{equation}
We seek to find a tortoise coordinate $r_*$ such that the coefficients of $dt^2$ and $dr_*^2$ are the same. This leads to the condition
\begin{equation}
\alpha dr_*=\frac{1}{Q}e^{-k_0 L-\frac{k_0^2}{2}W}dr.
\end{equation}
Using partial fraction decomposition, we can write $\frac{1}{Q}$ as
\begin{equation}
\frac{1}{(1-\alpha^2 r^2)(1-\frac{2m}{r})} = \frac{\beta_b}{1+\alpha r} + \frac{\beta_a}{1-\alpha r} + \frac{\beta_o}{1-\frac{2m}{r}}    
\end{equation}
where
\begin{equation}
\label{beta}
    \beta_b = \frac{1}{2(1+2\alpha m)}, \hspace{2em} \beta_a = -\frac{1}{2(1-2\alpha m)}, \hspace{2em} \beta_o = \frac{2\alpha m}{1-4\alpha^2 m^2}.
\end{equation}
Unlike in the original C-metric, this integral cannot be done exactly. However, the near-horizon leading order terms are the same as in the original C-metric for either horizon and hence, the definition of Kruskal coordinates will be of the same form.

\begin{figure}[ht]
\begin{subfigure}[t]{0.48\textwidth}
    \centering
    \includegraphics[height=4cm]{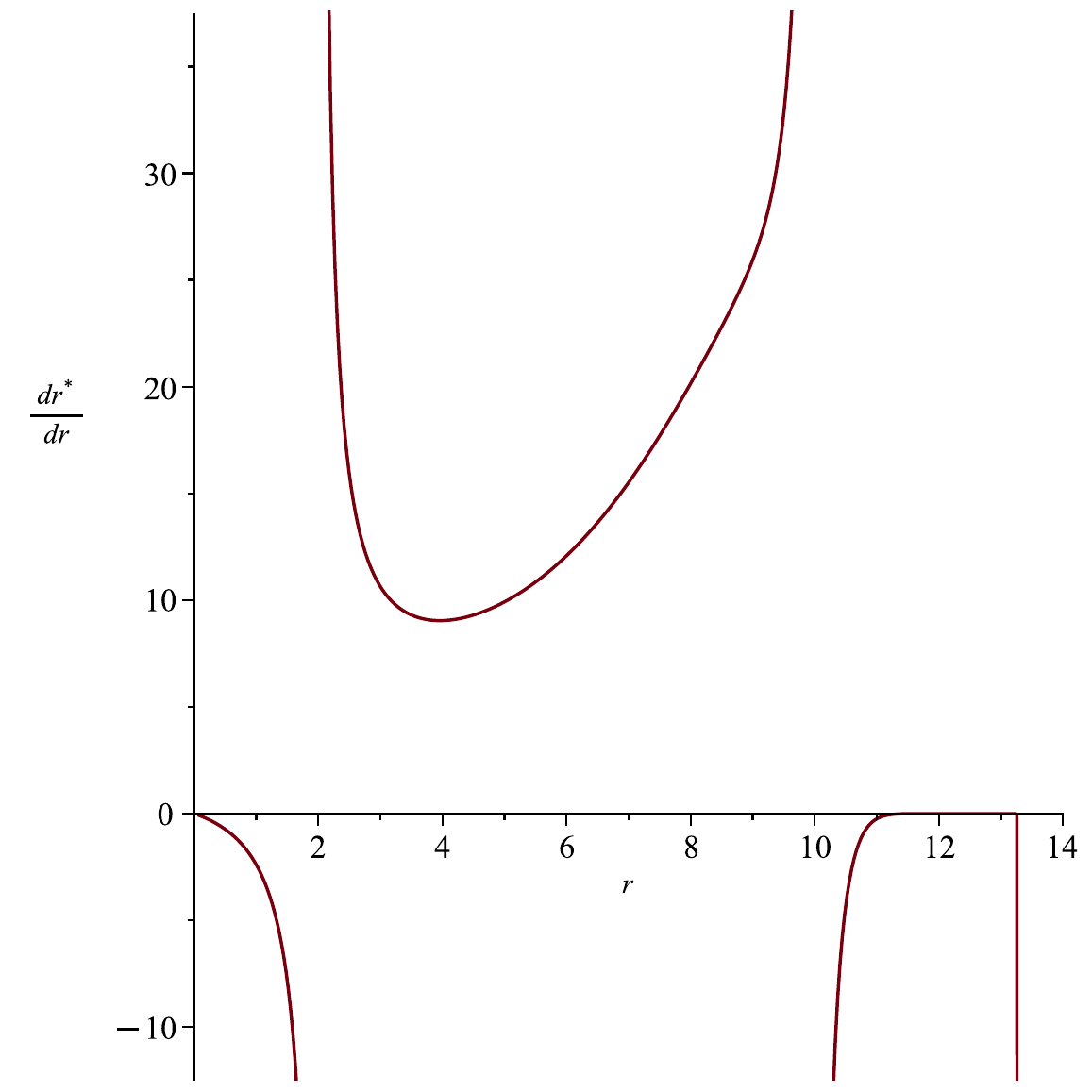}
    \caption{Plot of $\frac{dr_*}{dr}$ as a function of $r$}
    \label{fig:tortoise_coordinate_differential}
\end{subfigure}
\begin{subfigure}[t]{0.48\textwidth}
    \centering
    \includegraphics[height=4cm]{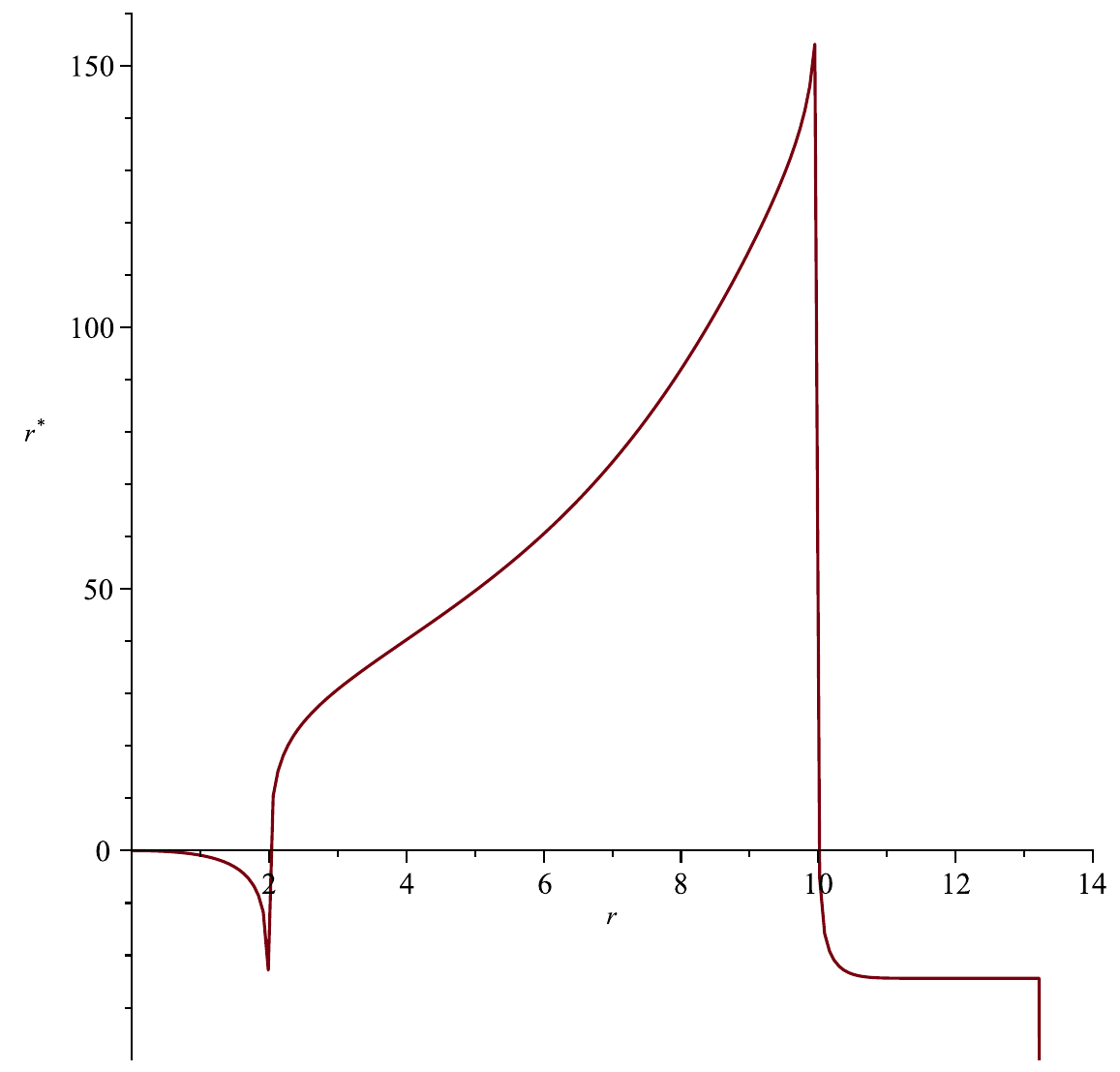}
    \caption{Plot of $r_*$ as a function of $r$}
    \label{fig:tortoise_coordinate}
\end{subfigure}
\caption{Visualizing the tortoise coordinate $r_*$ as a function of $r$, for $(m,\alpha)=(1,0.1)$.}
\end{figure}
With double-null coordinates 
\begin{equation}
    u=-r_*-t, \hspace{2em} v=-r_*+t
\end{equation}
the $2$-metric becomes
\begin{equation}
ds_2^2 = \frac{Qe^{k_0 (F+L)}}{(1-\alpha r \cos\theta)^2}(-du dv).
\end{equation}
We will analyze both horizons by modifying this form as needed.

\subsubsection*{Event Horizon $\mathcal{H}_o$}

\noindent To extend across $\mathcal{H}_o$, we define the Kruskal coordinates as follows\footnote{The factor $\text{sign}(\beta)$ is included so that in the overlapping region II, $U_o$ (resp. $V_o$) and $U_a$ (resp. $V_a$) will have opposite signs which ensures that they are oriented consistently with each other. The $i,j$ factors are included so that the formula is clearly invertible in the regions where $U,V$ are negative.}
\begin{equation}
    U_o = (-1)^i(-\text{sign}(\beta_o)) e^{-\frac{\alpha u}{2\beta_o}}, \hspace{2em} V_o = (-1)^j (-\text{sign}(\beta_o)) e^{-\frac{\alpha v}{2 \beta_o}},
\end{equation}
{ where $\beta_o$ is defined in equation (\ref{beta})}. The metric becomes
\begin{equation}
    ds_2^2 = -\frac{(-1)^{i+j}4\beta_o^2}{\alpha^2(1-\alpha r \cos\theta)^2 } Qe^{k_0 (F+L)} e^{-\frac{\alpha}{\beta_o} r_*} dU_o dV_o.
\end{equation}
It remains to be checked that the metric is indeed regular (finite and non-zero) across the horizon $r=2m$. Since $e^{k_0 (F+L)}$ is finite and non-zero at $r=2m$, it will not affect the calculation. Also, close to $r=2m$, $e^{-k_0 L-\frac{k_0^2}{2}W}$ is finite and non-zero, so it will not affect the behavior of the factor $\frac{1}{1-\frac{2m}{r}}$ in $\frac{dr_*}{dr}$. Hence, 
\begin{align}
\begin{split}
    \alpha r_* &\approx \beta_o \log|1-\frac{r}{2m}| \\
    \implies (1-\frac{2m}{r})e^{-\frac{\alpha}{\beta_o}r_*} &\approx \frac{2m}{r} \text{sign}\left(1-\frac{2m}{r}\right)
\end{split}
\end{align}
as with the original C-metric.

\subsubsection*{Acceleration Horizon $\mathcal{H}_a$}

\noindent Next, we define Kruskal-like coordinates to cross the horizon at $r=1/\alpha$ 
\begin{equation}
    U_a=(-1)^i (-\text{sign}(\beta_a))e^{-\frac{\alpha u}{2 \beta_a}}, \hspace{2em}
    V_a = (-1)^j (-\text{sign}(\beta_a)) e^{-\frac{\alpha v}{2 \beta_a}}.
\end{equation}
Then the metric becomes
\begin{equation}
    ds_2^2 = -\frac{4\beta_a^2}{\alpha^2} \frac{Q e^{k_0 (F+L)}}{(1-\alpha r \cos\theta)^2} e^{-\frac{\alpha}{\beta_a}r_*} dU dV.
\end{equation}
By the same argument as for the event horizon, the regularity of the acceleration horizon of the regularized C-metric follows from the regularity of the original C-metric.

\begin{figure}[htb]
    \centering
\begin{subfigure}[c]{0.45\textwidth}
\centering
    \includegraphics[height=5.5cm]{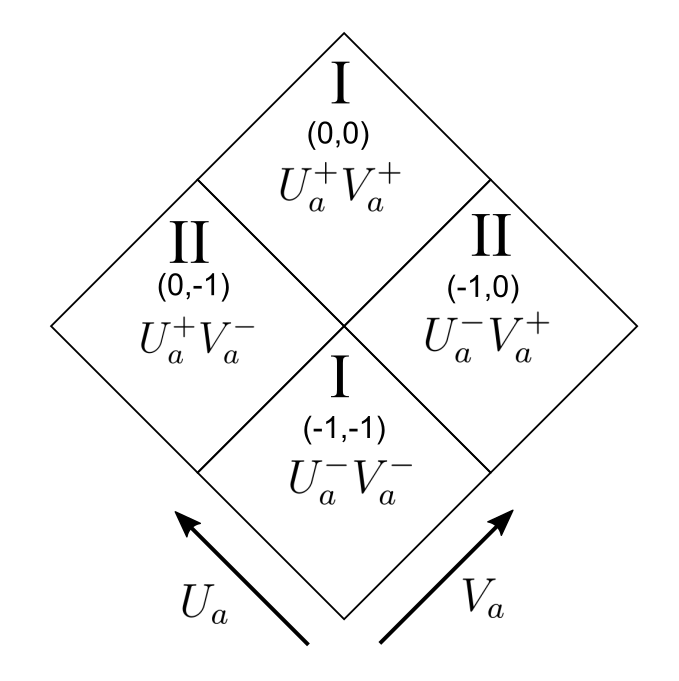}
    \caption{$\mathcal{H}_a$}
    \label{fig:cosmological}
\end{subfigure}
\begin{subfigure}[c]{0.45\textwidth}
\centering
    \includegraphics[height=5.5cm]{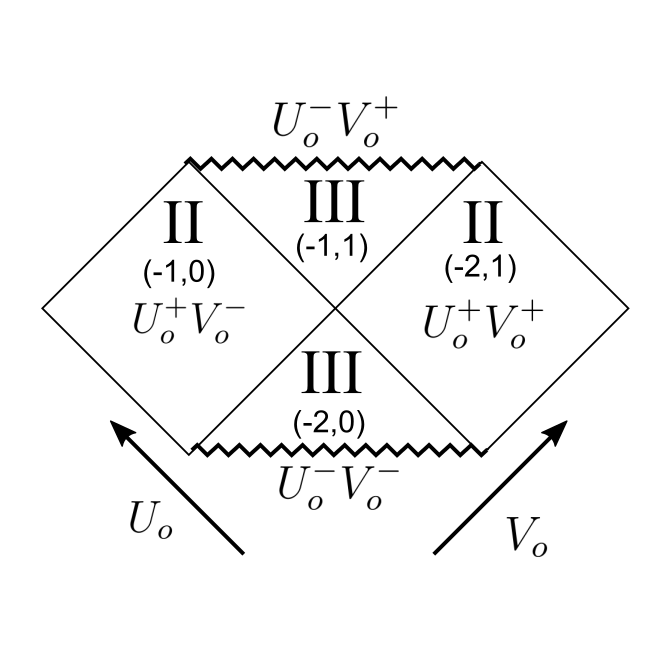}
    \caption{ $\mathcal{H}_o$}
    \label{fig:event}
\end{subfigure}
\caption{Diagrams showing Kruskal coordinates for the horizons $H_a$ and $H_o$ including the region type, the $(i,j)$ values, and signs of $U,V$. The regions of type II are a pair of regions exterior to a black hole. The regions of type III are a black hole and white hole connected by an Einstein-Rosen bridge. The regions of type I are the exterior of the acceleration horizon.}
\end{figure}

\subsubsection*{Identifying $\mathscr{I}$}

\noindent Finally, we compactify by defining
\begin{equation}
\tan\left(\frac{\tilde{u}}{2}\right) = U_a    \qquad \text{and} \qquad   \tan\left( \frac{\tilde{v}}{2} \right) = V_a
\end{equation}
which gives
\begin{equation}
    ds_2^2 = -\frac{\beta_a^2}{\alpha^2} \frac{Q e^{k_0 (F+L)}}{(1-\alpha r \cos\theta)^2} e^{-\frac{\alpha}{\beta_a}r_*} \sec^2\left(\frac{\tilde{u}}{2}\right)\sec^2\left(\frac{\tilde{v}}{2}\right)d\tilde{u} d\tilde{v}.
\end{equation}
To identify $\mathscr{I}$, we must identify where the metric coefficient diverges. The usual candidate would be $\tilde{u}=2\pi$ or $\tilde{v}=2\pi$, however, due to the denominator we must first check $r\rightarrow\frac{1}{\alpha \cos\theta}$ to see if the metric diverges even before reaching $\tilde{u}=2\pi$ or $\tilde{v}=2\pi$. Clearly the denominator goes to zero which seems like the metric component will become infinite. However, we must also investigate the other factors. In particular, we must check the behavior of the exponential factor $e^{k_0 (F+L)}$ which might approach either zero or infinity depending on the sign of $k_0 (F+L)$ near $r=\frac{1}{\alpha \cos\theta}$:
\begin{align}
    k_0 (F+L) &= \alpha k_0 \left(-\frac{r^2 \sin^2\theta (1-2\alpha m \cos\theta)}{(1-\alpha r \cos\theta)^2} + \frac{2m\cos\theta}{\alpha} + \frac{(1-\alpha^2 r^2)(1-\frac{2m}{r})}{\alpha^2 (1-\alpha r \cos\theta)^2} +\frac{2m}{\alpha^2 r}\right).
\end{align}
To check the sign, we combine the fractions and set $r=\frac{1}{\alpha \cos\theta}$ (ignoring the terms which do not diverge and the denominator which is positive):
\begin{align}
\begin{split}
    \text{numerator}&\equiv \alpha k_0 \left(-\alpha^2\frac{1}{\alpha^2 \cos^2\theta}\sin^2\theta (1-2\alpha m \cos\theta) + (1-\alpha^2 \frac{1}{\alpha^2 \cos^2\theta })(1-2\alpha m \cos\theta))\right) \\
    &=\alpha  k_0 \left(-\tan^2\theta(1-2\alpha m \cos\theta) -\tan^2\theta (1-2\alpha m \cos\theta) \right) \\
    &= -2\alpha k_0 \tan^2\theta (1-2\alpha m \cos\theta).
\end{split}
\end{align}
Since $k_0$ is negative for the value that regularizes the axis from equation \eqref{eqn:k-value}, the numerator is positive (since $2\alpha m <1$) so that the exponential overall approaches $+\infty$. Hence, $r=\frac{1}{\alpha \cos\theta}$ is the location of $\mathscr{I}$. We see that although the exact form of the metric has changed, the location of $\mathscr{I}$ is the same as in the case of the C metric.

\subsubsection*{Topology and metric of $\mathscr{I}$}
Although others discussed the radiative properties of the C-metric, these results were not made rigorous until the work of Ashtekar and Dray \cite{Ashtekar_Dray}. They emphasize how asymptotic flatness and admitting a construction of $\mathscr{I}$ is essential in discussing the presence of gravitational radiation. They proved that the C-metric is asymptotically flat at null infinity but not asymptotically Minkowskian \footnote{As mentioned in the introduction, one can regularize the nodal singularities of the charged C-metric by immersing it in an external electric or magnetic field and obtaining an electro-vac solution \cite{Ernst:1976}. The topology of $\mathscr{I}$ is again $\mathbb{R}\times \mathrm{S}^2$. However, in this case, one need not analytically extend the spacetime to get to this conclusion, since the existence of a conformal completion that is null, and whose derivative is everywhere regular on $\mathscr{I}$ implies that the topology is $\mathbb{R}\times \mathrm{S}^2$.}. One of the requirements for a spacetime to be asymptotically flat is that $\mathscr{I}$ has $\mathbb{R}\times \mathrm{S}^2$ topology\footnote{In \cite{Ashtekar_Dray}, the original statement of this condition is that the manifold of orbits of the restriction of $n^a$ to $\mathscr{I}$ should be diffeomorphic to $\mathrm{S}^2$. More recent literature states this condition as $\mathscr{I}$ having topology $\mathbb{R}\times \mathrm{S}^2$, see, for example, \cite{ashtekar2014}.}. To be asymptotically Minkowskian, a spacetime should satisfy the additional condition that the vector field $n^a:=\nabla^a\Omega$ generates all of $\mathscr{I}$, where $\Omega$ is the conformal factor used in the compactification. They showed that the C-metric fails this condition since there are two generators on $\mathrm{S}^2$ that only generate a half-line. For any analysis involving radiation, it is essential that the space of generators of $\mathscr{I}$ is topologically $\mathrm{S}^2$, therefore, if the topology of $\mathscr{I}$ were $\mathbb{R}^3$ (which would be the case if one whole generator were missing), such an analysis on the C-metric would have been impossible.

To rigorously construct $\mathscr{I}$ for the regularized C-metric, one would need a metric on $\mathscr{I}$ (like equation (3.4) in \cite{Ashtekar_Dray} for the original C-metric).
For this, one needs a suitable conformal factor, which is made difficult by the presence of exponential factors. In particular, one would need a conformal factor that dampens \emph{all} of the metric coefficients (see equation (\ref{C-metric-HongTeo})), which are infinite when $x+y=0$. The factor $\exp{\{k_0(F-L) - k_0^2W\}}$ which appears in front of the terms  containing $dx^2$ and $dy^2$ approaches infinity the fastest. Hence, we choose $\Omega = \alpha (x+y) \exp{\{-\frac{k_0}{2}(F-L)+\frac{k_0^2}{2}W\}}$. This satisfies the condition $\Omega= 0$ on $\mathscr{I}$, however, $\partial_\mu \Omega = 0$ and so this is not a suitable conformal factor. Additionally, this choice of $\Omega$ leads to a conformal metric that is entirely degenerate when $x+y=0$. Because the conformal  factor \emph{must} be of this form to dampen the blow up, it is not clear how one should proceed in finding a metric on $\mathscr{I}$.

For the regularized C-metric, the two points $\theta=0, \pi$ on $\mathrm{S}^2$ (at infinity) now contain singularities and must be excluded, so the submanifold becomes $\mathbb{R}\times \mathrm{S}$. 
Hence, in contrast to the missing half-generators in the case of the C-metric, the conformal infinity of the regularized C-metric is missing two whole lines (from $\mathbb{R}\times \mathrm{S}^2$) and thus has topology $\mathbb{R}^2\times \mathrm{S}$ which shows the generalized C-metric is not asymptotically flat. The topology of $\mathscr{I}$  can be made more rigorous by considering a maximal conformal extension in the sense of Chru\'{s}ciel \cite{chrusciel2010}, although this method would be somewhat less physical than the usual method of Penrose.

\subsubsection*{2D and 3D representation of the conformal diagrams}
\begin{figure}[htb]
\begin{subfigure}{0.32\textwidth}
\centering
\includegraphics[width=0.9\textwidth]{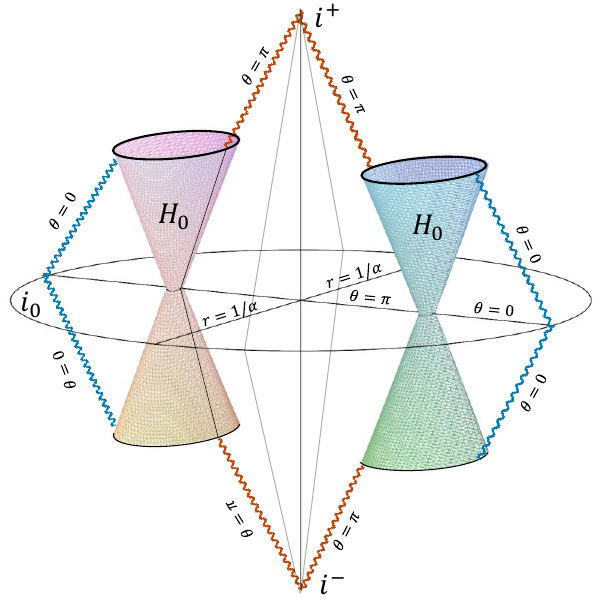}
\caption{Black hole horizons $H_0$}
\label{bh-3d}
\end{subfigure}\hfill
\begin{subfigure}{0.32\textwidth}
\centering
\includegraphics[width=0.9\textwidth]{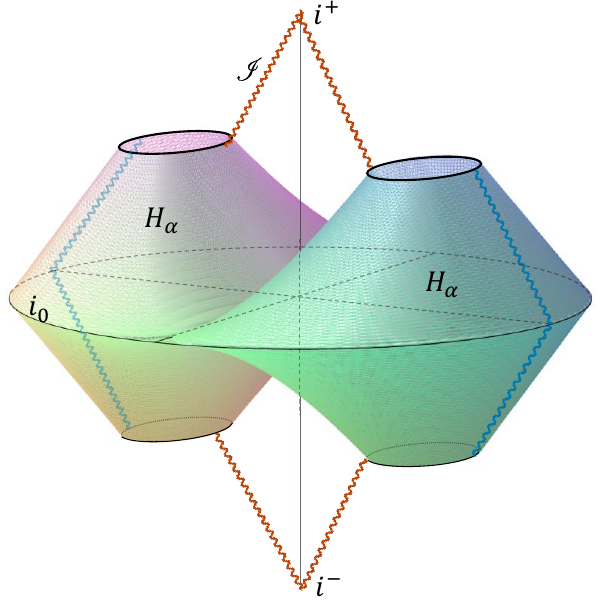}
\caption{Acceleration horizons $H_\alpha$}
\label{ah-3d}
\end{subfigure}\hfill
\begin{subfigure}{0.32\textwidth}
\centering
\includegraphics[width=0.9\textwidth]{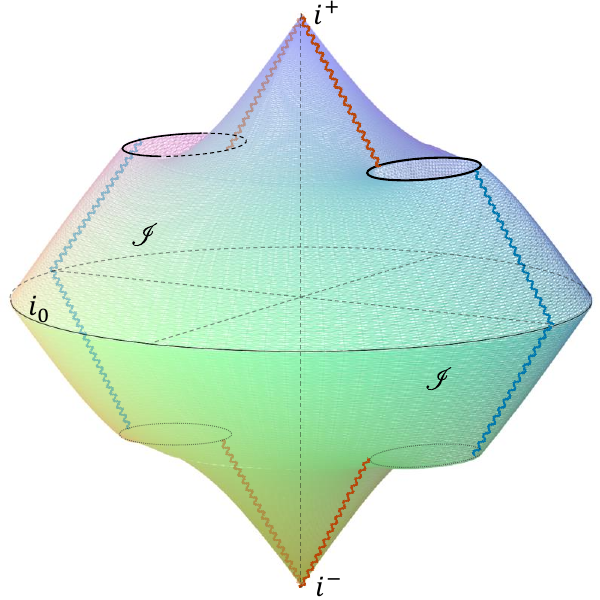}
\caption{Conformal infinity $\mathscr{I}$}
\label{scri-3d}
\end{subfigure}
\caption{ Figures \ref{bh-3d} and \ref{ah-3d} represent the two black hole horizons and the two acceleration horizons, respectively. Figure \ref{scri-3d} represents the conformal infinity. This shows clearly in 3D how these surfaces are nested inside each other, with $H_o$ contained inside $H_a$ which is contained inside $\mathscr{I}$. The dipole source term $kz$ (in Weyl coordinates) appears as singularities (corrugated lines) on $\theta=0$ and $\theta=\pi$ slices. It should be noted that a second copy of the coordinate $\theta$ is added to make the horizons (and other figures) circular. }
\label{3DFigures}
\end{figure}

In Figure \ref{2D-conformal-diagrams} the 2-dimensional conformal diagrams are presented for $\theta=0$, $\pi$ and a value in between. The dipole sources at infinity are indicated on the diagrams, which are colored to match the colors in Figure \ref{CoordPlot}. Since the spacetime is not spherically symmetric and, specifically, since $\mathscr{I}$ depends on the angle $\theta$, the 2D diagrams by themselves cannot serve as a representation of the global structure and hence, a 3D diagram is required. Figure \ref{3DFigures} shows the 3D representations of the black hole and acceleration horizons as well as conformal infinity $\mathscr{I}$.
These diagrams are created from the conformal pictures of the $m=0$ C-metric, that is, the 3D Minkowski conic diagrams. We start with the Minkowski coordinates as the base and write the C-metric's coordinates in terms of these, as in \cite{griffiths_krtous_podolsky}. A second copy of the coordinate $\theta$ is added to make the horizons look circular.

\begin{figure}[htb]
\begin{subfigure}[t]{0.18\textwidth}
\centering
\includegraphics[height=4cm]{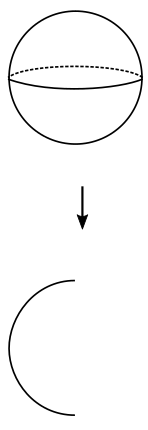}
\caption{$\mathrm{S}^2$ with $\phi$ suppressed}
\label{tent}
\end{subfigure}\hfill
\begin{subfigure}[t]{0.34\textwidth}
\centering
\includegraphics[height=4cm]{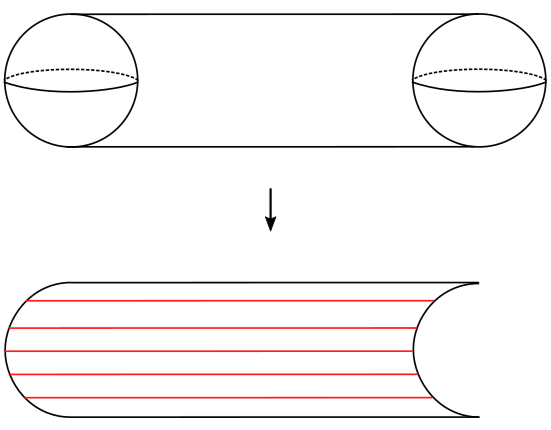}
\caption{$\mathbb{R}\times \mathrm{S}^2$ with $\phi$ suppressed}
\label{tent2}
\end{subfigure}\hfill
\begin{subfigure}[t]{0.46\textwidth}
\centering
\includegraphics[height=4.2cm]{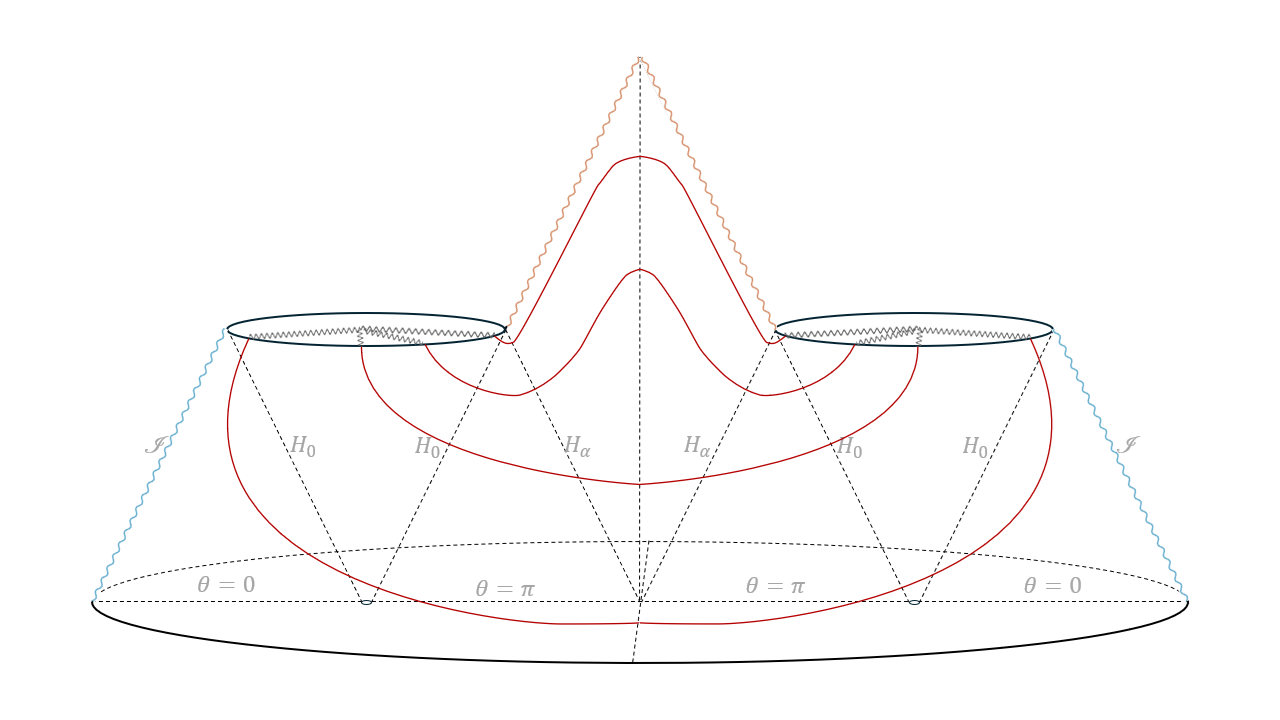}
\caption{Representing the different $\theta$ lines}
\label{tent3}
\end{subfigure}
\caption{ Fig \ref{tent2} shows the topological $ \mathbb{R}\times \mathrm{S}^2$ with azimuthal direction suppressed and the $\theta$-constant lines (in red). Figure \ref{tent3} shows the same lines warping as the ``half-cylinder'' is wrapped onto the ``carnival tent'' surface of $\mathscr{I}$. The half-cylinder is being wrapped onto the front side only. Similar to the usual diamond diagram for Minkowski which includes a superfluous ``mirror image'' for aesthetic purposes, in these 3D diagrams, only the front half of the diagram is necessary.}
\label{tent-diagram}
\end{figure}

By construction, the 2D diagrams can be obtained as a $\theta$-slice of the 3D diagram. The appearance of $\mathscr{I}$ as a spacelike line in the 2D conformal diagrams may seem strange. Because these diagrams are a 2D section of the 3D diagram, the 2D representation of $\mathscr{I}$ is given by the intersection of the full $\mathscr{I}$ with a spacelike hypersurface $\theta=const.$  The 3D diagram shows only one pair of black holes due to pictorial limitations (we cannot show the static region on the ``other side'' of the black hole), but the maximal extension will still contain an infinite sequence of black holes. Figure \ref{tent-diagram} shows how the different $\theta$-constant lines wrap around the 3-dimensional $\mathscr{I}$.

\subsubsection*{Other $k$ values}

\noindent
For any $k\ne \frac{1}{4m} \ln\frac{1-2\alpha m}{1+2\alpha m}$ the metric will be conically singular. However, the previous analysis shows that for any $k\leq 0$, $\mathscr{I}$ would be the same as for the regularized case. For values $k>0$, the exponential would reach $0$ and would cause the metric to become degenerate at $\mathscr{I}$. Since this calculation is based on a 2-dimensional metric, it is not immediately clear if $\mathscr{I}$ would become completely degenerate, i.e., a point, or if it is only $\theta$-slices which are degenerate. The inability to find a full 4-dimensional coordinate system adapted to the null structure prevents one from reaching a conclusion.

\subsubsection*{Wormhole Interpretation}
\begin{figure}[tb]
\begin{subfigure}[t]{0.48\textwidth}
    \centering
    \includegraphics[height=4.4cm]{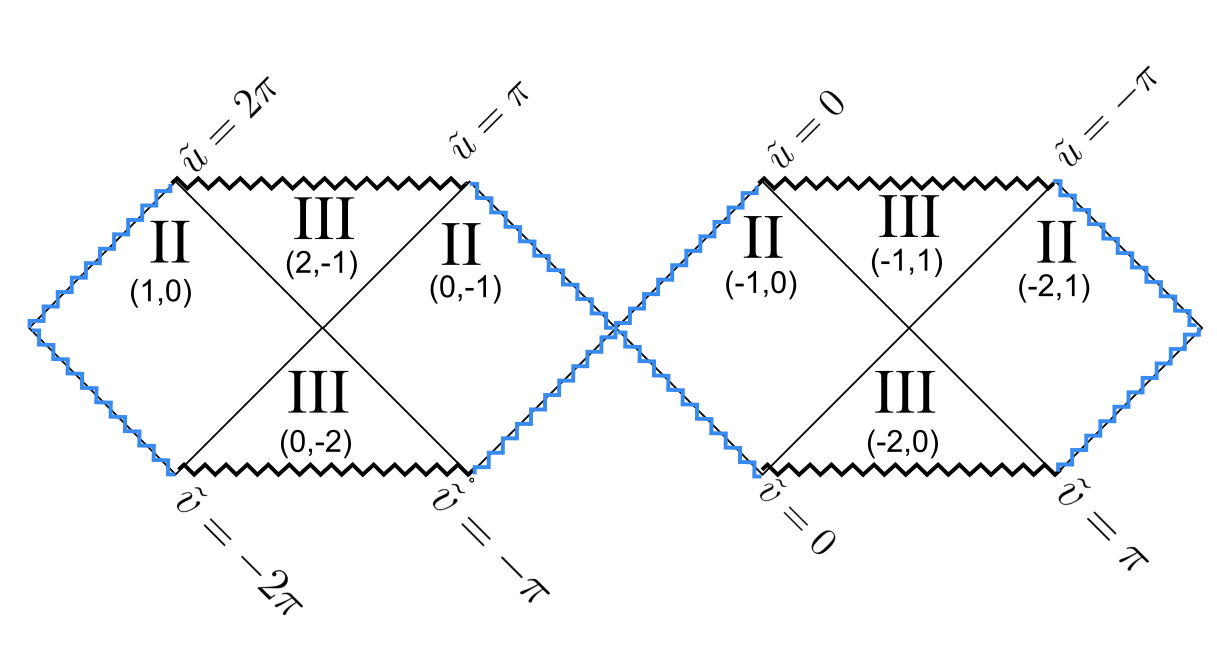}
    \caption{Section of conformal diagram for $\theta=0$ which has a curvature singularity at ${r=1/\alpha}$.}
    \label{thetaZero}
\end{subfigure}
\hfill
\begin{subfigure}[t]{0.48\textwidth}
    \centering
    \includegraphics[height=4.4cm]{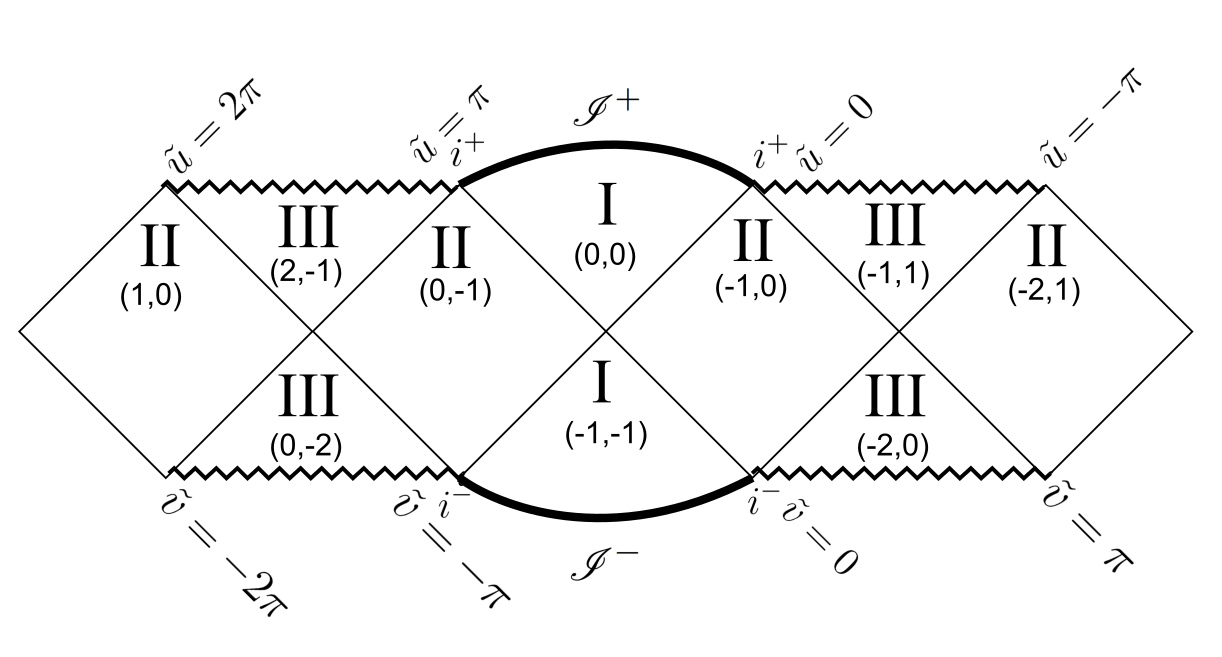}
    \caption{Section of conformal diagram for ${0<\theta<\pi}$. For these sections, there is no curvature singularity and $\mathscr{I}$ is located at ${r=1/\alpha\cos\theta}$.}
    \label{thetaMid}
\end{subfigure}
\begin{center}
\begin{subfigure}[t]{0.6\textwidth}
    \centering
    \includegraphics[height=4.4cm]{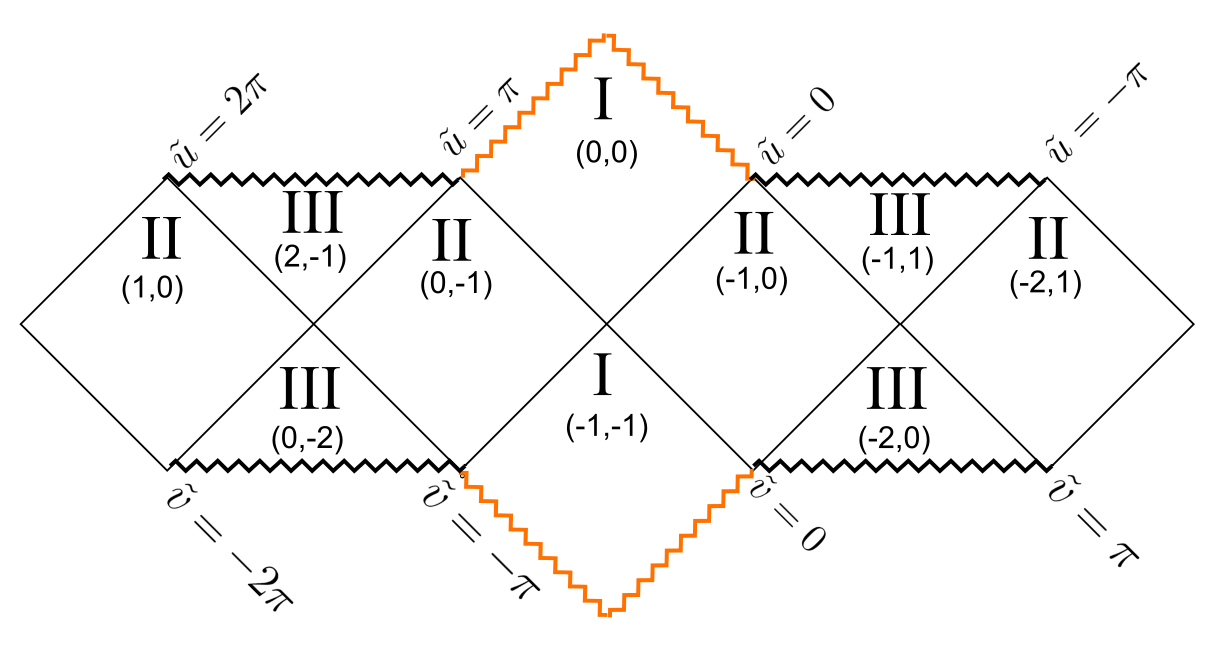}
    \caption{Section of conformal diagram for $\theta=\pi$. In this case, there is another curvature singularity that cannot be reached by the $r$ coordinate. In terms of $x,y$ coordinates, this is the section $x=-1$ and has curvature singularity at $y=1$.}
    \label{thetaPi}
\end{subfigure}
\end{center}
\caption{Although spacelike infinity $i^0$ appears in the 3D conformal diagram, it does not appear in any of the $\theta$ sections. This is due to the fact that restricting to a $\theta$ section means spacelike lines will traverse across different static regions without end. In order to reach $i^0$, a path must cut through different $\theta$ sections and hence this cannot be seen in a 2D section. For the same reason, $i^+$ and $i^-$ do not appear in regions of type I. In (b), the line representing $\mathscr{I}$ is not manifestly null as it is a $\theta$ slice (spacelike) of the full hypersurface $\mathscr{I}$ which is null, as seen in \ref{3DFigures}. The original C-metric has only the $r=0$ singularities, but the behavior of $i^0, i^+$, $i^-$, and $\mathscr{I}$ is the same. Going from Figure \ref{thetaPi} to \ref{thetaZero}, one can see that $\mathscr{I}$ ``bows down'' as $\theta$ decreases, as in the case of C-metric \cite{Kinnersley_Walker}. (For the original C-metric, Figure 11 in \cite{Kinnersley_Walker} shows the conformal diagram for the C-metric as a single diagram by superposing the different $\theta$ slices.)} 
\label{2D-conformal-diagrams}
\end{figure}

As noted by \cite{Kinnersley_Walker} for the original C-metric, one may still choose a nontrivial topology for the regularized C-metric. This leads to a (non-traversable) wormhole which can be visualized from Figure \ref{2D-conformal-diagrams} by identifying the regions $(2,-1)$ and $(-1,1)$, for example.

\subsection{Bondi Coordinates?}

The generalized C-metric also does not admit standard Bondi coordinates. For the generalized C-metric,
\begin{equation}
e^{-4u+2\nu} = \frac{\left[ (1-2\alpha m)R_1 + (1+2\alpha m)R_2 + 4\alpha m R_3 \right]^2}{8\alpha \left(1-4\alpha^2 m^2\right)^2 R_1 R_2 R_3}   \frac{R_1+R_2+2m}{R_1+R_2-2m} \left( R_3 + z + \frac{1}{2\alpha} \right)^{-1} \frac{e^{2kF-k^2\rho^2-4kz}}{\alpha}~
\end{equation}
which has the asymptotic expansion
\begin{equation}
{\mathrm e}^{ {k \cos\Theta \left(4m- \frac{1}{\alpha}\right)} + k R \left(\cos\Theta + 1\right) \left(kR\cos\Theta -kR - 2\right)} \Biggl[ \frac{\left(\cos\Theta+1\right)^{-1}}{2 \alpha^{2} \eta^{2} R^{2}}  +  \frac{8 m \alpha^{2}+k \cos\Theta-2 \alpha -k}{8 \alpha^{4} \eta^2 R^{3}}  + \mathcal{O}\left(\frac{1}{R^4}\right) \Biggr]~,
\end{equation}
where $\eta = 1 - 2\alpha m$. The presence of the exponential factor prevents further computations.

\subsubsection*{Nonstandard Bondi Coordinates}
To attempt to find nonstandard Bondi coordinates, a natural first step is to find a null coordinate such as in equation (\ref{c-metric-null}). This is clearly not possible due to the $\theta$ dependence of the metric functions. One might be tempted to try a more general coordinate transformation such as, for example, a transformation of the form
\begin{equation}
    u=u(t,r,\theta), \hspace{2em} R=R(r,\theta).
\end{equation}
In this case, the constraints for Bondi coordinates (eliminating all $dR$ terms except $dudR$) leads to a contradiction and cannot be satisfied. A more general transformation attempt would be
\begin{equation}
    u=u(t,r,\theta), \hspace{2em} R=R(r,\theta),\hspace{2em} \mu=\mu(r,\theta).
\end{equation}
In this case, the constraints for the Bondi coordinates are highly nonlinear and intractable. Explicitly, the constraint equations are as follows:
\begin{align}
    \frac{g_{tt}}{u_{,t}^2}\left( u_{,\theta}^2 \mu_{r,}^2 -2 u_{r,} u_{,\theta} \mu_{,r} \mu_{,\theta} + u_{,r}^2 \mu_{,\theta}^2 \right) + g_{rr} \mu_{,\theta}^2 + g_{\theta \theta} \mu_{,r}^2 = 0 \\
    \frac{g_{tt}}{u_{,t}^2}\left( u_{,\theta} \mu_{,r} - u_{,r} \mu_{,\theta}  \right)\left( u_{,r} R_{,\theta} - u_{,\theta} R_{,r} \right) - g_{rr}\mu_{,\theta} R_{,\theta} - g_{\theta \theta } \mu_{,r} R_{,r} = 0
\end{align}
where $g_{ij}$ are the metric coefficients in the $(t,r,\theta, \phi)$ coordinates.

\section{Conclusion}
{  The C-metrics are intriguing exact solutions in general relativity that describe accelerating black holes in static coordinates.} Following its interpretation as a pair of accelerating black holes, the original C-metric and its various matter generalizations have been studied from different angles in both classical and quantum gravity. In particular, a detailed understanding of the maximally extended spacetime exists for both the vacuum and charged C-metrics. One problem with C-metrics, in general, is the existence of conical singularities on their axes. For the (original) charged C-metric, this can be removed by introducing an electric field of appropriate strength \cite{Ernst:1976}. However, astrophysical black holes are sparsely charged and do not seem to possess conical singularities. Hence, a purely gravitational way of regularizing the vacuum C-metric could provide a concrete model of axially symmetric static black holes that are accelerating. As mentioned in the Introduction, there is potentially an infinite number of ways to regularize the vacuum C-metric using the inverse scattering method, of which Ernst's method is the simplest one, which introduces a dipole source. However, the inverse scattering method requires the use of Weyl coordinates, and one loses the advantage provided by Hong-Teo and other coordinate systems all of which are amenable to Ernst's prescription.

The generalized C-metric, a one-parameter extension of the C-metric in pure gravity, received little attention despite Ernst's initial enthusiasm. The reason remains unclear --- this may be due to the complexity of the metric functions or uncertainty about the correct parameter value for regularity. The goal of this paper was to extend the analyses performed on C-metrics over the years to the regularized C-metric{ , and, as a result, bring it to similar levels of understanding as the original C-metric.} To achieve this, we first had to resolve the disagreement in the literature regarding which parameter value makes the metric non-singular on the axis. In this connection, we analyzed the conical singularities of the C-metric employing the generalized Gauss-Bonnet theorem, which we believe represents the first application of this theorem to  C-metrics. This allowed us to examine carefully the nuances involved in the conical singularities of the original as well as regularized C-metrics and to study their embeddability. We also analyzed Bonnor's conjecture that the C-metric can be written in the standard Bondi form starting from Weyl coordinates, and found that this is not the case.

The generalized C-metric fails to be asymptotically flat under the same analysis of Ashtekar and Dray that established that the original C-metric is asymptotically flat. The analysis offers a precise geometric reason for non-flatness: the presence of curvature singularities at the poles of $\mathrm{S}^2$ necessitates the deletion of lines from conformal infinity. In any coordinate system, the regularized C-metric requires much more careful consideration than the original C-metric, due to the complication that arises from the additional exponential factors. Despite this, we found that the regularized C-metric can be analytically extended beyond the horizons leading to a maximal extension similar to that of the original C-metric. We constructed 2D and 3D conformal diagrams, as done in \cite{griffiths_krtous_podolsky} for the vacuum C-metric. To aid our analysis, we introduced additional visual tools --- the ``informal diagram" and the ``carnival tent" --- which we believe will provide useful perspectives for understanding other axisymmetric spacetimes. For the regularized C-metric, we have found that one cannot rigorously construct $\mathscr{I}$. However, the 2D and 3D conformal diagrams still capture the causal structure of the spacetime. The  two black holes accelerating away from each other without requiring a line source confirms Ernst’s hypothesis that conical singularities arise due to the absence of an explicit physical source explaining the acceleration. 

To conclude, we have revisited a vacuum solution in general relativity that has lain dormant for several decades, despite its ability to describe a pair of black holes in vacuum without conical singularities. We have explored the underlying spacetime and analyzed its classical properties in sufficient detail to bring it on par with the original C-metric. With this work, we hope that this spacetime is now accessible enough to become a subject of further research.

\section*{Acknowledgements}
The authors would like to thank Ji\v{r}\'{i} Podolsk\'{y} for useful communications.  A special thanks to Mat\'u\v{s} Papaj\v{c}\'ik for careful reading of the manuscript and for many useful comments and suggestions. We also thank the anonymous referees for useful suggestions.

\appendix

\section{Connecting with Dray and Walker} \label{Dray-Walker-connection}
In Section \ref{ernst-c-metric}, we presented the generalized C-metric obtained by applying Ernst's method. Here, we briefly compare the work of Dray and Walker with ours. Equation $12a$ of Dray and Walker computes the external field strength to be
\begin{equation}
\lambda = \frac{1}{2MA(x_1-x_2)}\left(\ln\frac{|x_1(1+3MAx_1)|}{|x_2(1+3MAx_2)|}\right)
\end{equation}
which, after appropriate transformations, does not match our results. The auxiliary fields computed by Dray-Walker contain a small error. Their $A$ should be replaced by $1/A$ for the generalized metric to be a vacuum solution, that is, the field strengths should read
\begin{equation}
L_{DW} = -(r^2F + 2My/A)~, \qquad N_{DW} = r^2G + 2Mx/A~.
\end{equation}
instead of $L_{DW} = -(r^2F + 2MAy)$ and $N_{DW} = r^2G + 2MAx$. This error affects the subsequent analysis of the Ernst external field strength, which should be
\begin{equation}
\lambda = \frac{A}{2M(x_1-x_2)}\left(\ln\frac{|x_1(1+3MAx_1)|}{|x_2(1+3MAx_2)|}\right)~.
\end{equation}
Once this correction is made, the following rescalings must be performed to match their results with ours. The constants rescale as
\begin{equation}
A = \frac{\alpha}{B}~, \qquad M = \frac{m}{c_0^3}, \qquad x_1 = Bc_0 (1 - c_1)~, \qquad x_2 = Bc_0 (-1 - c_1)~,
\end{equation}
where
\begin{equation}
\frac{1}{B^2} = 1 + c_1^2 + 4 \alpha m c_1^3~, \qquad c_0^2 = 1+ 6\alpha m c_1~, \qquad c_1 = \frac{\sqrt{1+12m^2\alpha^2}-1}{6\alpha m}~.
\end{equation}
In addition, since the exponential terms should match, $\lambda$ and the auxiliary fields transform as
\begin{equation}
(L_{DW}, N_{DW}) \to \frac{B^2}{c_0^2}(-L, -F)~, \qquad  \lambda = -\frac{c_0^2}{B^2}k~,
\end{equation}
where the negative signs are merely a matter of convention. Using all these,
\begin{equation}
\lambda = \frac{c_0^2}{B^2} \frac{\alpha}{4m} \ln \frac{1 + 2\alpha m}{1 - 2\alpha m} = - \frac{c_0^2}{B^2} k~,
\end{equation}
as expected. For completeness, $\tilde{t}$ and $\tilde{\phi}$ re-scale as $\tilde{t}= \frac{c_0}{B}t$, $\tilde{\phi}= \frac{c_0}{B} \varphi$.

\section{Relation with metric coefficients}\label{top_s2}
In reference to the discussion in Section \ref{gb}, we derive here the condition metric coefficients satisfy for the surface to be $\mathrm{S}^2$. Let $M$ be a two-dimensional manifold with an axially symmetric metric
\begin{equation}\label{2-metric}
ds^2 = f(\theta) d\theta^2 + \sin^2\theta d\phi^2~,
\end{equation}
where $\theta\in [0,\pi]$. Let this be the submanifold of a four-dimensional Lorentzian manifold with $\partial_{\phi}$ as one of the Killing vectors, and we would like this submanifold to be of spherical topology. The roots of $\sin^2\theta$ determine the axis of the spacetime, that is $\theta\in \{0,\pi\}$, and thus the regularity must be checked around these two points. The Gaussian curvature for such spaces is given by
\begin{equation}\label{gen-gaussian}
K = -\frac{1}{2\sin\theta\sqrt{f}}\left( \frac{\partial}{\partial \theta} \frac{2\cos\theta}{\sqrt{f}} \right)~,
\end{equation}
and
\begin{equation}
\int_M K dA = 2\pi C\left[ \frac{-\cos\theta}{\sqrt{f}}\right]_{0}^{\pi} = 2\pi C\left[ \frac{1}{\sqrt{f}|_0} + \frac{1}{\sqrt{f}|_\pi} \right]
\end{equation}
where $2 \pi C$ is the period of $\phi$. To check for conical singularities, one needs to compute the ratio of circumference to radius around the axis. This gives
\begin{equation}
\lim_{\theta\to \{0,\pi\}} \frac{\int \sin\theta d\phi}{\int \sqrt{f} d\theta} = \frac{2 \pi C}{\sqrt{f}|_{\{0,\pi\}}}~.
\end{equation}
Hence, for such spaces to be two-spheres without conical singularities,
\begin{equation}\label{sphere-condition-0}
\underbrace{2\pi C\left[ \frac{1}{\sqrt{f}|_0} + \frac{1}{\sqrt{f}|_\pi} \right]}_{\int K~dA} + \underbrace{\frac{2 \pi C}{\sqrt{f}|_0} + \frac{2 \pi C}{\sqrt{f}|_\pi} - 4\pi}_{\sum \beta_i} = 4\pi~,
\end{equation}
implying
\begin{equation}
C\left[ \frac{1}{\sqrt{f}|_0} + \frac{1}{\sqrt{f}|_\pi} \right] = 2~.
\end{equation}
It is straightforward to show that a conformal factor on the metric in \eqref{2-metric} does not change the result as long as the conformal factor is smooth and non-zero for $\theta \in [0,\pi]$.

\section{Curvature Scalars for the generalized C-metric}\label{sec:w1w2}
The following Maple code has been used to compute the curvature scalars for the generalized C-metric in Hong-Teo coordinates \eqref{C-metric-HongTeo}--\eqref{C-metric-HongTeo-func}:
\begin{lstlisting}[basicstyle=\footnotesize]
    restart;
    with(DifferentialGeometry): with(Tensor):
    DGsetup([t, y, x, phi], C) ; 
    con := (x, y) -> alpha^2*(x + y)^2 ;
    A := y -> -(1 - y^2)*(1 - 2*alpha*m*y) ;
    B := x -> (1 - x^2)*(1 + 2*alpha*m*x) ;
    F := (x, y) -> -B(x)/con(x, y) - 2*m*x/alpha ;
    L := (x, y) -> A(y)/con(x, y) + 2*m*y/alpha ;
    g := evalDG(-A(y)*exp(k*(F(x, y) + L(x, y)))*(dt &t dt)/con(x, y) + B(x)*exp(-k*(F(x, y) + L(x, y)))*(dphi &t dphi)/con(x, y) + exp(k*(F(x, y) - L(x, y)) - k^2*A(y)*B(x)/con(x, y)^2)*(dy &t dy/A(y) + dx &t dx/B(x))/con(x, y)) ;
    G := EinsteinTensor(g)
                        G := 0 D_t D_t
    RI := RiemannInvariants(g, author = "CarminatiMcLenaghan");
\end{lstlisting}
The (two) nonvanishing curvature scalars ($w_1$ and $w_2$) are:
{\tiny

}

\printbibliography[heading=bibintoc, title={References}]

\end{document}